\documentclass[a4,11pt]{article}

\usepackage{amsfonts,latexsym,eucal,color,graphicx,epsfig,amssymb,amsmath,bm}
\usepackage{color}

\newcommand{\be}{\begin{eqnarray}}
\newcommand{\ee}{\end{eqnarray}}
\newcommand{\beq}{\begin{eqnarray}}
\newcommand{\eeq}{\end{eqnarray}}
\newcommand{\pd}{\partial}

\newcommand{\nn}{\nonumber}

\newcommand{\dalm}{\kern1pt\vbox{\hrule height 0.9pt\hbox{\vrule width 0.9pt\hskip 2.5pt\vbox{\vskip 5.5pt}\hskip 3pt\vrule width 0.3pt}\hrule height 0.3pt}\kern1pt}

\makeatletter

\@addtoreset{equation}{section}
\makeatletter

\begin{document}

\thispagestyle{empty}
\begin{titlepage}

\begin{flushright}
\end{flushright}
\vspace{2cm}
\begin{center}
{\Large {\bf Vacuum excitation by sudden appearance and\\
\vspace{2mm}
disappearance of a Dirichlet wall in a cavity}} \\
\vspace{15mm}

{\bf Tomohiro Harada${}^{1}$ \quad Shunichiro Kinoshita${}^{2}$ \quad Umpei Miyamoto${}^{3}$ }\\

\vspace{10mm}

{\small {\it
${}^{1}$Department of Physics, Rikkyo University, Tokyo 171-8501, Japan\\
}}
{\small {\tt
harada@rikkyo.ac.jp \\
}}

\vspace{5mm}

{\small {\it
${}^{2}$Department of Physics, Faculty of Science and Engineering, Chuo University,\\
1-13-27 Kasuga, Bunkyo-ku,Tokyo 112-8551, Japan \\
}}
{\small {\tt 
kinoshita@phys.chuo-u.ac.jp\\
}}

\vspace{5mm}
{\small {\it
${}^{3}$RECCS, Akita Prefectural University, Akita 015-0055, Japan \\
}}
{\small {\tt
umpei@akita-pu.ac.jp \\
}}

\end{center}

\vspace{5mm}

\begin{abstract}
Vacuum excitation by time-varying boundary conditions is not only of fundamental importance but also has recently been confirmed in a laboratory experiment. In this paper, we study the vacuum excitation of a scalar field by the instantaneous appearance and disappearance of a both-sided Dirichlet wall in the middle of a 1D cavity, as toy models of bifurcating and merging spacetimes, respectively. It is shown that the energy flux emitted positively diverges on the null lines emanating from the appearance and disappearance events, which is analogous to the result of Anderson and DeWitt. This result suggests that the semiclassical effect prevents the spacetime both from bifurcating and merging. In addition, we argue that the diverging flux in the disappearance case plays an interesting role to compensate for the lowness of ambient energy density after the disappearance, which is lower than the zero-point level.
\end{abstract}
\end{titlepage}

\newpage
\section{Introduction}
\label{sec:intro}

There is general belief that quantum effects dominate in a final stage of some physical phenomena to avoid undesired results predicted by classical theories. In the context of gravitational physics, one may say that the chronology protection conjecture is a manifestation of such belief stated in a verifiable form~\cite{Hawking:1991nk}. This conjecture asserts that the emergence of a closed timelike curve, which can be thought of as a natural time machine and lead to paradoxes, would be prevented by the backreaction of explosive particle creation.  

A wormhole of spacetime has many interesting properties but is unfavorable in the sense that its existence leads to the emergence of closed timelike curves~\cite{Morris:1988tu}. It is not surprising if semiclassical effects act in the direction of preventing the dynamical formation of wormholes, even if it is allowed in classical frameworks.

\begin{figure}[b]
\begin{center}
\begin{minipage}[c]{0.8\textwidth}
\linespread{1}
\begin{center}
\includegraphics[height=3.5cm]{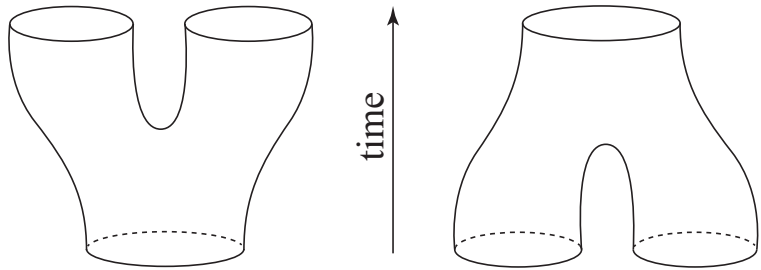}
\caption{{Schematic pictures of spacetime splitting (left) and 
merger (right). Quantum fields living in these spacetimes undergo the 
drastic change of environments such as boundary conditions.
}}
\label{fg:tr}
\end{center}
\end{minipage}
\end{center}
\end{figure}

The formation of a wormhole can happen in the merger of disconnected two spaces (see the right panel of Fig.~\ref{fg:tr}, and see Ref.~\cite{Maeda:2008bh} for a simple exact solution to the Einstein equation representing such a wormhole formation). Quantum fields living in such a spacetime will undergo the drastic change of their environments. For example, the spacetime merger gives rise to the sudden increase of spatial volume and the sudden change of boundary conditions for the quantum fields near the merger point. What we are concerned in this paper is whether or not these sudden changes excite the vacuum of quantum fields and the backreactions to the spacetime play crucial roles.

Particle creation in the spacetime merger has not been studied extensively in the literature, as far as the present authors know (but see \cite{Braunstein:1996aj}). On the other hand, the particle creation in some situations analogous to the spacetime splitting (see the left panel of Fig.~\ref{fg:tr}) has been investigated in the literature. A pioneering work is that of Anderson and DeWitt~\cite{Anderson:1986ww}. They considered the dynamical change of spatial topology, $S^1 \to S^1 + S^1$ (consider the cut of the boundary of splitting spacetime in Fig.~\ref{fg:tr} by constant-time planes), and the vacuum excitation of a test scalar field in such a background. They found that an explosive flux of created particles is emitted from the `crotch', suggesting that the backreaction prevents the spacetime from bifurcating. This conclusion was supported by~\cite{Manogue}.

In this paper, we consider a test scalar field in a 1D Dirichlet
cavity. We assume that a both-sided Dirichlet wall can appear and
disappear instantaneously at the center of cavity, which mimics the
sudden splitting and merger of spacetime, respectively.
As we will see, when the central wall suddenly appears, the initial vacuum is highly
excited to result in a strong flux in an almost same way as the
Anderson-DeWitt analysis. Namely, the flux contains a delta function squared multiplied by a logarithmically diverging factor. On the other hand, when the central wall suddenly disappears, the initial vacuum is also highly excited to result in a diverging flux, although it does not contain the delta function squared.

As we will see, the total number of created particles diverges both in
the appearance and disappearance cases. This means that two ground
states defined when the Dirichlet wall is absent and present are
orthogonal. Namely, the vacuum structures are completely different
before and after the appearance and disappearance. 
Therefore, we expect that the diverging flux appears in the both cases. 
In fact, we will see that this is the case. We will discuss this point in Conclusion again.

Before starting our analysis, we note that the vacuum excitation by time-varying boundary conditions, as those considered in this paper, is not only interesting from the gravitational physics point of view, but also one of hot topics in the fundamental studies of relativistic quantum field theory. As predicted by Moore in 1970~\cite{Moore}, the non-inertial motion of cavity boundary induces the emission of photons, which is called the dynamic Casimir effect. Although the rapid acceleration of boundary, of which speed has to be comparable with the speed of light, was experimental challenge, the boundary motion turned out to be effectively realized by modulating the electromagnetic properties of boundary with high frequencies. Then, the particle creation was recently observed in a laboratory experiment using superconducting circuit~\cite{nature}. It is beyond the scope of the present paper to list the references before or after this experimental breakthrough. See, e.g. \cite{Brown:2015yma}, and references therein for recent developments and updated interests such as the quantum information theory and black-hole firewall problem.

The organization of this paper is as follows. In Sec.~\ref{sec:app}, we investigate the vacuum excitation due to the sudden appearance of the both-sided Dirichlet boundary. While the obtained result in this section is up to our expectations, we can prepare for the succeeding investigations and test the validity of our formulation. In Sec.~\ref{sec:disapp}, we investigate the vacuum excitation due to the sudden disappearance of the Dirichlet boundary. Most parts of the analysis proceed in parallel with those in Sec.~\ref{sec:app}, but the results are different. In Sec.~\ref{sec:smooth}, we discuss the vacuum excitation by the smooth appearance and disappearance of Dirichlet wall adopting the formulation in Ref.~\cite{Brown:2015yma}, and compare its instantaneous limit with the results in Secs.~\ref{sec:app} and \ref{sec:disapp}. Section~\ref{sec:conc} is devoted to a conclusion. The proof and derivation of several equations are given in Appendices. In particular, the vacuum expectation values of energy-momentum tensor in both appearance and disappearance cases, which are main results of this paper, are reproduced in Appendix~\ref{sec:green} using the Green functions, rather than the Bogoliubov transformation used in the text. We use the natural unit in which $c=\hbar=1$.

\section{Sudden appearance of a Dirichlet boundary}
\label{sec:app}

\subsection{Classical behaviors of a massless scalar field}

\begin{figure}[bht]
\begin{center}
\begin{minipage}[c]{0.8\textwidth}
\linespread{1}
\begin{center}
\includegraphics[height=6cm]{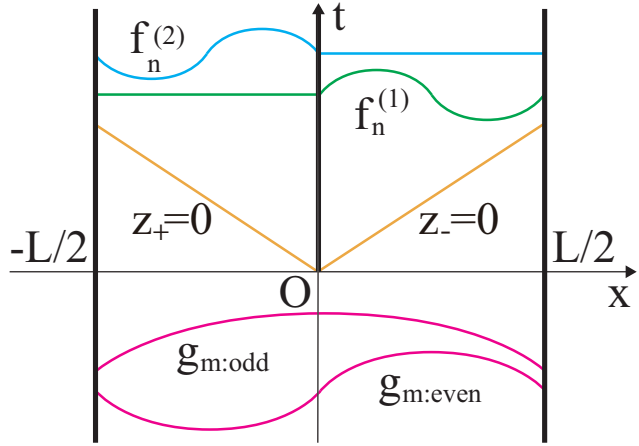}
\caption{The sudden appearance of a Dirichlet boundary in a 1D
 cavity. The scalar field obeys the Dirichlet boundary conditions at the
 both ends $(x=\pm \frac{L}{2})$ for $-\infty<t<\infty$ and at the
 center $(x=0)$ for $t>0$. The null lines $z_\pm := t \pm x = 0 \; (t>0)$ and the spatial configurations of mode
 functions $g_m$ and $f_n^{(\gamma)} \; (m,n \in {\bf N},
 \gamma \in \{1,2 \})$ are schematically depicted.
}
\label{fg:bc1}
\end{center}
\end{minipage}
\end{center}
\end{figure}

We consider a massless scalar field confined in a 1D cavity, which obeys the following equation of motion
\be
	(-\pd_t^2 + \pd_x^2 ) \phi (t,x)=0,
\;\;\;
	-\infty<t<\infty,
\;\;\;
	-\frac{L}{2} \leq x \leq \frac{L}{2}.
\label{eom}
\ee
We impose the Dirichlet boundary conditions at the both ends
\be
	\phi (t,\pm \frac{L}{2}) = 0,
\;\;\;
	-\infty<t<\infty.
\label{BC_end}
\ee
In addition, we impose another Dirichlet boundary condition at the center after $t=0$,
\be
	\phi (t, 0) = 0,
\;\;\;
	t > 0,
\label{BC_App}
\ee
in order to model the sudden appearance of a perfect mirror in the cavity (see Fig.~\ref{fg:bc1}).

Preparing for the quantization of the scalar field, we shall find appropriate positive-energy mode functions before and after $t=0$.

In the past asymptotic region $t \to -\infty$, the following $\{
g_m \} \; ( m \in {\bf N} )$ constitute a set of positive-energy mode functions,
\be
	g_m(t,x)
	=
	\frac{1}{\sqrt{m\pi}} e^{ -i q_m t} \times
	\begin{cases}
		\cos q_m x  & (m: \mbox{odd}) \\
		\sin  q_m x  & (m: \mbox{even})
	\end{cases},
\;\;\;
	q_m := \frac{m\pi}{L},
\;\;\;
	-\frac{L}{2} \leq x \leq \frac{L}{2},
\label{g}
\ee
which satisfy the orthonormal conditions
\be
	\langle g_m, g_{m'} \rangle =- \langle g_m^\ast, g_{m'}^\ast \rangle = \delta_{mm'},
\;\;\;
	\langle g_m,g_{m'}^\ast \rangle =0.
\ee
Here, the asterisk denotes the complex conjugate and $ \langle \phi,\psi \rangle :=i\int_{-L/2}^{L/2} (\phi^\ast \pd_t \psi - \pd_t \phi^\ast \psi) dx $ is the Klein-Gordon inner product, evaluated on a spacelike curve  $t={\rm const.}$~\cite{Birrell:1982ix}. This inner product is conserved, namely independent of time, whenever both $\phi$ and $\psi$ are solutions to the equation of motion~\eqref{eom} and vanish on every boundary. Note that the above expression of $g_m$, Eq.~\eqref{g}, is valid only before the appearance of the Dirichlet wall.

In the future asymptotic region $t \to +\infty$, the following $\{ 
f_n^{(\gamma)} \}\; (\gamma \in \{1,2\},\;  n \in {\bf N})$ constitute a set of positive-energy mode functions,
\begin{align}
\begin{split}
&
	f^{(1)}_n (t,x)
	=
	\begin{cases}
		\displaystyle 0 & \displaystyle ( -\frac{L}{2} \leq x < 0 )\\
		\displaystyle \frac{1}{\sqrt{n\pi}} e^{-i p_n t} \sin p_n x & \displaystyle ( 0 \leq x \leq \frac{L}{2} )
	\end{cases},
\\
&
	f^{(2)}_n (t,x)
	=
	\begin{cases}
		\displaystyle -\frac{1}{\sqrt{n\pi}} e^{ -i p_n t } \sin p_n x
		&  \displaystyle ( -\frac{L}{2} \leq x < 0 ) \\
		\displaystyle 0 &  \displaystyle ( 0 \leq x \leq \frac{L}{2} )
	\end{cases},
\;\;\;
	p_n := \frac{2n\pi}{L},
\label{f}
\end{split}
\end{align}
which satisfy the orthonormal conditions
\be
	\langle f_{n}^{(\gamma)}, f_{n'}^{(\gamma')} \rangle
	=
	- \langle f_{n}^{(\gamma)\ast}, f_{n'}^{(\gamma')\ast} \rangle
	=
	\delta_{\gamma\gamma'} \delta_{nn'},
\;\;\;
	\langle f_{n}^{(\gamma)}, f_{n'}^{(\gamma')\ast} \rangle = 0.
\ee
Noted that the above expression of $f_m^{(\gamma)}$, Eq.~\eqref{f}, is valid only after the appearance of the Dirichlet wall.

We consider the expansion of $g_m$ by $f_n^{(\gamma)}$,
\be
	g_m
	=
	\sum_{\gamma=1}^2 \sum_{n=1}^\infty
	(
		\rho_{mn}^{(\gamma)} f_n^{(\gamma)}
		+
		\sigma_{mn}^{(\gamma)} f_n^{(\gamma)\ast}
	),
\label{gBYf}
\ee
where the expansion coefficients, called the Bogoliubov coefficients, are evaluated as
\be
	\rho_{mn}^{(\gamma)}
	=
	\langle f_n^{(\gamma)}, g_m \rangle,
\;\;\;
	\sigma_{mn}^{(\gamma)}
	=
	- \langle f_n^{(\gamma)\ast}, g_m \rangle.
\label{bogo}
\ee
We take $t=0$ as the spacelike curve on which the inner products in Eq.~\eqref{bogo} are evaluated. Substituting Eqs.~\eqref{g} and \eqref{f} into Eq.~\eqref{bogo}, we obtain
\be
	\rho_{mn}^{(\gamma)}
	=
	\begin{cases}
		\displaystyle \frac{2}{(2n-m)\pi} \sqrt{ \frac{n}{m} }  & (m:\mbox{odd}) \\
		\displaystyle \frac{ (-1)^{\gamma-1} }{\sqrt{2}}\delta_{m,2n} & (m:\mbox{even})
	\end{cases},
\;\;\;
	\sigma_{mn}^{(\gamma)}
	=
	\begin{cases}
		\displaystyle \frac{2}{(2n+m)\pi} \sqrt{ \frac{n}{m} }  & (m:\mbox{odd}) \\
		\displaystyle 0 & (m:\mbox{even})
	\end{cases}.
\label{rho_sigma}
\ee

Here, we note that the expansion \eqref{gBYf} for odd $m$ is valid
everywhere except for $x=0$  (namely, almost everywhere in a mathematical sense). 
This is because the mode functions $g_m$ for odd $m$ can take non-zero values at $x=0$ while the mode functions $f_n$ are always zero by the boundary conditions. We will look into the implication of Eq.~\eqref{gBYf} in Conclusion, comparing with corresponding relation~\eqref{f_g} in the disappearance case.

\subsection{Quantization of the scalar field}

The canonical quantization of the scalar field is implemented by expanding the field operator $ {\bm \phi} $ by two set of mode functions, $\{ g_m \}$ or $\{ f_n^{(\gamma)} \}$, as
\begin{align}
	{\bm \phi}
	&=
	\sum_{m=1}^\infty
	(
		{\bm b}_m g_m + {\bm b}_m^\dagger g_m^\ast	
	)
\label{phi_g}
\\
	&=
	\sum_{\gamma=1}^2 \sum_{n=1}^\infty
	(
		{\bm a}_n^{(\gamma)} f_n^{(\gamma)}
		+
		{\bm a}_n^{(\gamma) \dagger } f_n^{(\gamma) \ast } 
	),
\label{phi_f}
\end{align}
and imposing the commutation relations on the expansion coefficients
\begin{align}
	[ {\bm b}_m, {\bm b}_{m'}^{\dagger} ] 
	=
	\delta_{mm'},
\;\;\;
&
	[ {\bm b}_m, {\bm b}_{m'} ] 
	=
	0,
\label{CR_b}
\\
	[ {\bm a}_n^{(\gamma)}, {\bm a}_{n'}^{(\gamma')\dagger} ] 
	=
	\delta_{\gamma \gamma'} \delta_{nn'},
\;\;\;
&
	[ {\bm a}_n^{(\gamma)}, {\bm a}_{n'}^{(\gamma')} ] 
	=
	0.
\label{CR_a}
\end{align}
Then, ${\bm b}_m$ and ${\bm a}_n^{ (\gamma) }$ are interpreted as annihilation operators, and ${\bm b}_m^\dagger$ and ${\bm a}_n^{(\gamma) \dagger }$ creation operators. 

Substituting Eq.~\eqref{gBYf} into Eq.~\eqref{phi_g}, and comparing it with Eq.~\eqref{phi_f}, one obtains
\be
	{\bm a}_n^{(\gamma)}
	=
	\sum_{m=1}^\infty
	(
		\rho_{mn}^{(\gamma)} {\bm b}_m
		+
		\sigma_{mn}^{(\gamma) \ast} {\bm b}_m^\dagger
	).
\label{aBYb}
\ee
Substituting Eq.~\eqref{aBYb} into Eq.~\eqref{CR_a}, and using Eq.~\eqref{CR_b}, one finds that the following conditions must hold for the two quantizations, Eqs.~\eqref{phi_g} and \eqref{phi_f}, to be consistent.  
\begin{align}
	\sum_{m=1}^\infty
	(
		\rho_{mn}^{(\gamma)} \rho_{mn'}^{(\gamma') \ast}
		-
		\sigma_{mn}^{(\gamma)\ast} \sigma_{mn'}^{(\gamma')}
	)
	&=
	\delta_{\gamma \gamma'} \delta_{nn'},
\label{Consis1}
\\
	\sum_{m=1}^\infty
	(
		\rho_{mn}^{(\gamma)} \sigma_{mn'}^{(\gamma') \ast}
		-
		\sigma_{mn}^{(\gamma)\ast} \rho_{mn'}^{(\gamma')}
	)
	&=
	0.
\label{Consis2}
\end{align}
Hereafter, we call these relations unitarity relations. In Appendix~\ref{sec:con1}, we prove that the Bogoliubov coefficients given by Eq.~\eqref{rho_sigma} satisfy these unitarity relations.

Since we are interested in the particle creation caused by the appearance of boundary, we assume that the quantum field is in the vacuum state $| 0_g \rangle$ in which any particle corresponding to $g_m$ does not exist. Such a vacuum is characterized by
\be
	{\bm b}_m | 0_g \rangle =0,
\;\;\;
	\langle 0_g | 0_g \rangle =1,
\;\;\;
	\forall m \in {\bf N}.
\label{vac1}
\ee

\subsection{Spectrum and energy-momentum density}

The vacuum $| 0_g \rangle$ contains no particle corresponding to $g_m$ but can contain particles corresponding to $f_n^{(\gamma)}$. This is examined by calculating the vacuum expectation value of particle-number operator,
\begin{align}
	\langle 0_g | {\bm a}_n^{(\gamma)\dagger} {\bm a}_n^{(\gamma)} | 0_g \rangle
	=
	\sum_{m=1}^\infty | \sigma_{mn}^{(\gamma)} |^2
	=
	\frac{4n}{\pi^2}
	\sum_{\substack{ m=1 \\ m:{\rm odd}}}^\infty
	\frac{1}{ m (m+2n)^2 }.
\label{NumDen2}
\end{align}
While this is finite, its summation over $n $ and $\gamma $, i.e.\ the total number of created particles, diverges.
This implies that the Fock space representation associated with ${\bm
b}_{m}$ is unitarily inequivalent to that associated with ${\bm a}_n^{(\gamma)}$~\cite{Wald:1995yp,Rodriguez-Vazquez:2014hka}.

In order to see directly what happens, we calculate the vacuum
expectation value of the energy-momentum tensorial operator, which is
given for the massless scalar field by $ {\bm T}_{\mu\nu} = \pd_\mu {\bm
\phi} \pd_\nu {\bm \phi} - \frac12 \eta_{\mu\nu} (\pd {\bm
\phi})^2$. Here, $\eta_{\mu\nu}={\rm Diag}.\ (-1,1)$ is the two-dimensional
Minkowski metric. If one introduces the double null coordinates, the non-zero components of the energy-momentum operator are
\be
	{\bm T}_{\pm \pm} = ( \pd_\pm {\bm \phi} )^2,
\;\;\;
	z_{\pm} := t \pm x.
\label{tpm}
\ee
Note that the energy density and momentum density (or energy-flux density) in the original Cartesian coordinates are given by $  {\bm T}^{tt} = {\bm T}_{--} + {\bm T}_{++}$ and ${\bm T}^{tx} = {\bm T}_{--} - {\bm T}_{++}$, respectively.

Substituting Eq.~\eqref{phi_g} into Eq.~\eqref{tpm}, and using
Eq.~\eqref{g}, we obtain the vacuum expectation value before the
appearance of the Dirichlet boundary,
\begin{align}
	\langle 0_g | {\bm T}_{\pm\pm} | 0_g \rangle_{t<0}
	=
	\sum_{m=1}^\infty | \pd_\pm g_m |^2
	=
	\frac{\pi}{4L^2} \sum_{m=1}^\infty m.
\label{Tab_02}
\end{align}
This is clearly divergent but can be renormalized by 
standard procedures~\cite{Birrell:1982ix}.
If we remove the ultraviolet divergence caused by the vacuum energy of
the Minkowski space (namely $L\to\infty$), then we obtain well-known
finite result as 
\begin{equation}
\langle 0_g | {\bm T}_{\pm\pm} | 0_g \rangle_{{\rm ren}, t<0} = -
 \frac{\pi}{48L^2}. 
\label{eq:simple_Casimir}
\end{equation}
Such a negative energy is called the Casimir energy. In the ordinary 3D electromagnetic case, this kind of negative energy gives rise to an attractive force between two parallel neutral plates put in vacuum~\cite{Casimir}.

What we are most interested in is the same quantity after the appearance
of the Dirichlet boundary. Substituting Eq.~\eqref{phi_f} into Eq.~\eqref{tpm}, and then using Eq.~\eqref{aBYb}, we obtain
\begin{align}
&
	\langle 0_g| {\bm T}_{\pm\pm} |0_g \rangle_{t>0}
\nn
\\
&=
	\sum_{\gamma=1}^2 \sum_{ \substack{ m=1 \\ m:{\rm odd} } }^\infty \sum_{n=1}^\infty \sum_{n'=1}^\infty 
	[
		( \rho_{mn}^{(\gamma)} \sigma_{mn'}^{(\gamma)} + \rho_{mn'}^{(\gamma)} \sigma_{mn}^{(\gamma)} )
		{\rm Re}
		( \pd_\pm f_n^{(\gamma)} \pd_\pm f_{n'}^{(\gamma)} )
		+
		( \rho_{mn}^{(\gamma)} \rho_{mn'}^{(\gamma)} +  \sigma_{mn}^{(\gamma)} \sigma_{mn'}^{(\gamma)} )
		{\rm Re}
		( \pd_\pm f_n^{(\gamma)} \pd_\pm f_{n'}^{(\gamma)\ast} )  
	]
\nn
\\
&+
	\sum_{\gamma=1}^2 \sum_{ \substack{ m=2 \\ m: {\rm even} } }^\infty \sum_{n=1}^\infty \sum_{n'=1}^\infty 
	\rho_{mn}^{(\gamma)} \rho_{mn'}^{(\gamma)} {\rm Re}
	( \pd_\pm f_n^{(\gamma)} \pd_\pm f_{n'}^{(\gamma)\ast} ).
\label{TAB_01}
\end{align}
To derive Eq.~\eqref{TAB_01}, we symmetrize the dummy indices $n$ and $n'$. In addition, we use the facts that $ \sigma_{mn}^{(\gamma)} $ vanishes for even $m$, and $\pd_\pm f_n^{(1)}$ and $\pd_\pm f_{n'}^{(2)}$ have no common support, i.e.\ $ \pd_\pm f_n^{(\gamma)} \pd_\pm f_{n'}^{(\gamma')} \propto \delta_{\gamma \gamma'}$.

Substituting explicit form of Bogoliubov coefficients~\eqref{rho_sigma} and mode functions~\eqref{f} into Eq.~\eqref{TAB_01}, we obtain
\begin{align}
&
	\langle 0_g| {\bm T}_{\pm\pm} |0_g \rangle_{t>0}
\nn
\\
&=
	\frac{1}{\pi L^2} \sum_{ \substack{ m=1 \\ m:{\rm odd} } }^\infty
	\left(
		\frac{1}{4m}
		[ 
			4 \sum_{n=1}^\infty  \cos ( \frac{2n\pi}{L} z_\pm )
			+
			m^2 \sum_{n=1}^\infty \frac{ \cos ( \frac{2n\pi}{L} z_\pm ) }{ n^2-(m/2)^2 }
		]^2
		+
		m
		[
			\sum_{n=1}^\infty \frac{ n \sin ( \frac{ 2n\pi }{L} z_\pm ) }{ n^2-(m/2)^2 } 
		]^2
	\right)
+
	\frac{\pi}{4L^2} \sum_{ \substack{ m=2 \\ m: {\rm even} } }^\infty m.
\label{TAB_02}
\end{align}
This is an even function of $z_\pm$ with period $L$, as it is invariant
under reflection $z_\pm \to - z_\pm$ and translation $z_\pm \to z_\pm +
L$. Therefore, it is sufficient to calculate it in $0 \leq z_\pm < L$,
and then generalize the obtained expression to the one valid in the entire domain appropriately.

The first and second summations over $n$ in Eq.~\eqref{TAB_02} can be calculated to give
\begin{align}
&
	\langle 0_g| {\bm T}_{\pm\pm} |0_g \rangle_{t>0}
\nn
\\
&=
	\frac{1}{\pi L^2} \sum_{ \substack{ m=1 \\ m:{\rm odd} } }^\infty
	\left(
		\frac{1}{4m}
		[ 
			4 L^2 \delta (z_\pm)^2 + m^2 \pi^2 \sin^2 ( \frac{m\pi}{L} z_\pm )
		]
		+
		m
		[
			\sum_{n=1}^\infty \frac{ n \sin ( \frac{ 2n\pi }{L} z_\pm ) }{ n^2-(m/2)^2 } 
		]^2
	\right)
+
	\frac{\pi}{4L^2} \sum_{ \substack{ m=2 \\ m: {\rm even} } }^\infty m,
\label{TAB_03}
\end{align}
which is valid in $ 0 \leq z_\pm < L $. Here, we have used the following formulas,
\begin{align}
	\sum_{k=1}^\infty \cos( \frac{ 2k \pi}{a} y )
	&=
	-\frac12 + \frac{a}{2} \sum_{\ell = -\infty}^\infty \delta( y- \ell a ),
\;\;\;
	( - \infty  < y < \infty ),
\\
	\sum_{k=1}^\infty
		\frac{ \cos ky }{ k^2-a^2 }
		&=
		-\frac{\pi}{2a} \cos[ a(\pi - y) ] {\rm cosec} (a \pi )+\frac{1}{2a^2},
\;\;\;
	(0 \leq y \leq 2\pi),
\end{align}
where $\delta$ represents the Dirac delta function. See Ref.~\cite[p.~730]{maru} for the second formula.

For $z_\pm = 0$, from Eq.~\eqref{TAB_03}, we have
\be
	\langle 0_g| {\bm T}_{\pm\pm} |0_g \rangle_{t>0}
	=
	\frac{ 1 }{\pi}
	\sum_{ \substack{ m=1 \\ m:{\rm odd} } }^\infty
	\frac{ \delta(0)^2 }{m}
	+
	\frac{\pi}{4L^2} \sum_{ \substack{ m=2 \\ m: {\rm even} } }^\infty m,
\;\;\;
	(z_\pm = 0).
\label{TAB_04}
\ee
For $ 0 < z_\pm < L$, the summation over $n$ in Eq.~\eqref{TAB_03} can be calculated to give
\be
	\langle 0_g| {\bm T}_{\pm\pm} |0_g \rangle_{t>0}
	=
	\frac{\pi}{4L^2} \sum_{m=1}^\infty m,
\;\;\;
	(0<z_\pm < L),
\label{TAB_05}
\ee
using the following formula~\cite[p.~730]{maru}
\begin{align}
	\sum_{k=1}^\infty
	\frac{ k \sin ky }{ k^2-a^2 }
	&=
	\frac{\pi}{2} \sin[ a(\pi - y) ] {\rm cosec} (a \pi ),
\;\;\;
	(0 <y < 2\pi).
\end{align}

Extending the domain of Eqs.~\eqref{TAB_04} and \eqref{TAB_05} to the entire domain periodically, we obtain 
\be
	\langle 0_g| {\bm T}_{\pm\pm} |0_g \rangle_{t>0}
	=
	\frac{  1 }{\pi}
	\sum_{ \substack{ m=1 \\ m:{\rm odd} } }^\infty
	\frac{ 1 }{m}
	\sum_{\ell=-\infty}^\infty \delta( z_\pm - \ell L )^2
	+
	\begin{cases}
	\displaystyle
	\frac{\pi}{4L^2} \sum_{ \substack{ m=2 \\ m: {\rm even} } }^\infty m
	&
	(z_\pm = \ell L, \; \ell \in {\bf Z} ) \\
	\displaystyle
	\frac{\pi}{4L^2} \sum_{m=1}^\infty m
	&
	({\rm otherwise})
	\end{cases}.
\label{Tab_03}
\ee

\begin{figure}[bht]
\begin{center}
\begin{minipage}[c]{0.8\textwidth}
\linespread{1}
\begin{center}
		\setlength{\tabcolsep}{ 0 pt }
		\begin{tabular}{ cc }
			\includegraphics[height=6.5cm]{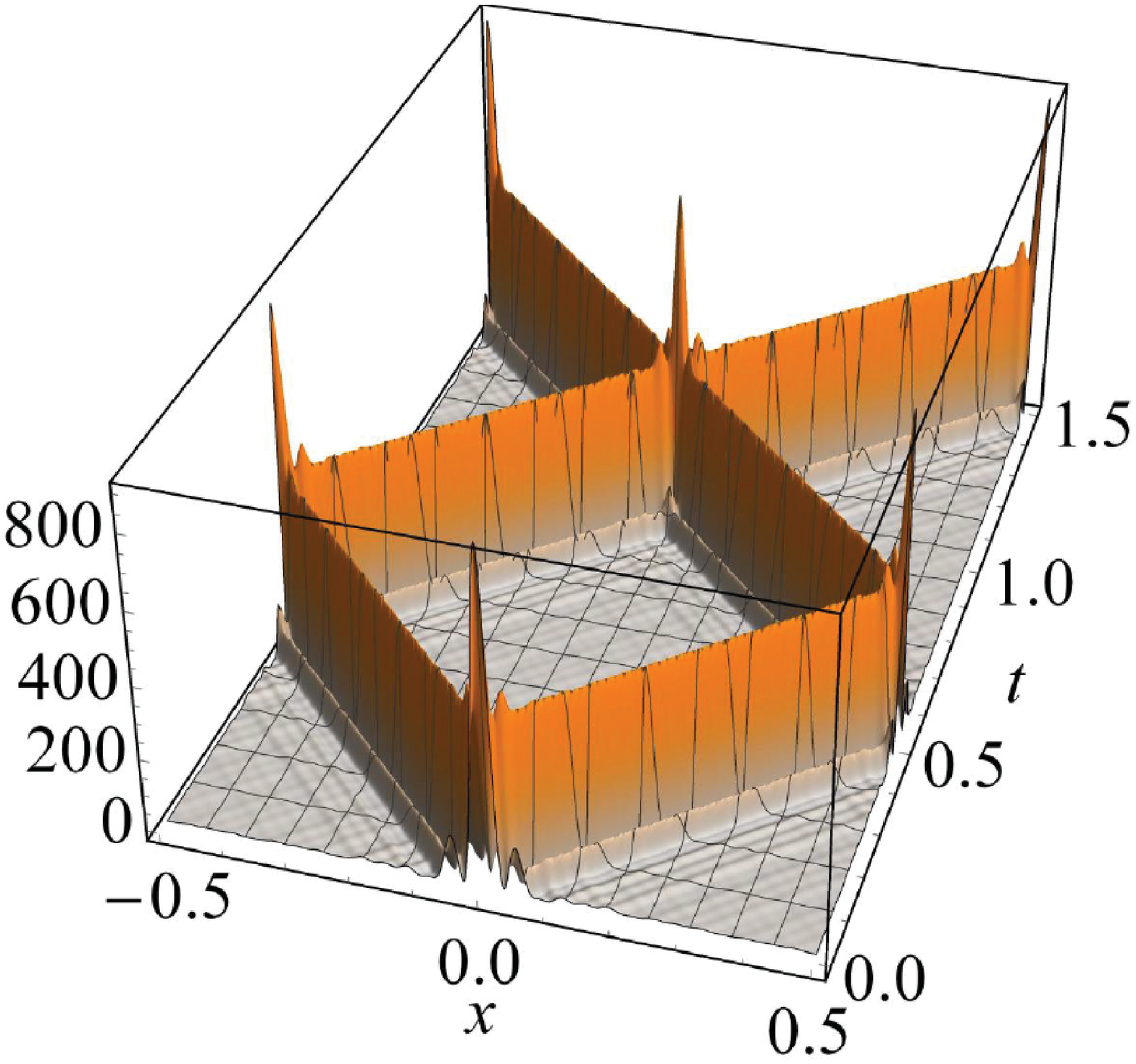} &
			\includegraphics[height=6.5cm]{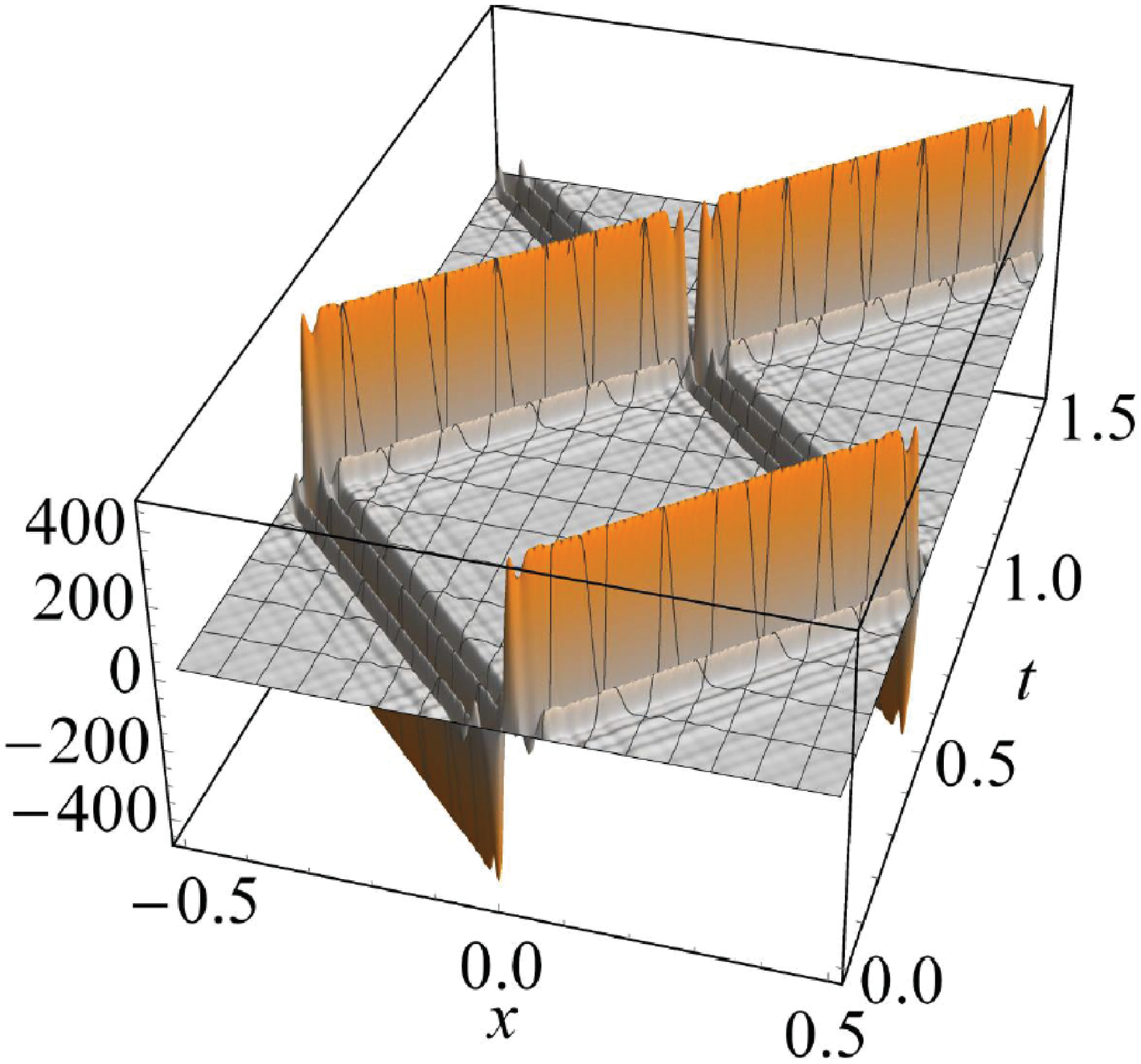} \\
		\end{tabular}
\caption{The vacuum expectation values of energy density $\langle 0_g | ( {\bm T}_{--} + {\bm T}_{++} ) | 0_g \rangle_{t>0}$ (left) and momentum density $ \langle 0_g | ({\bm T}_{--} - {\bm T}_{++}) | 0_g \rangle_{t>0} $ (right) with cutoff from which the Casimir contribution is subtracted. We set $L=1$ and summation over modes in Eq.~\eqref{TAB_02} is taken until $n = m = 14$. The exact results without cutoff are given by Eq.~\eqref{Tab_03}.
}
\label{fg:VEV}
\end{center}
\end{minipage}
\end{center}
\end{figure}

From the above result, we immediately see that an infinitely strong energy flux which
behaves as the delta function squared multiplied by a logarithmically
divergent factor emanates from the appearance point of the Dirichlet
boundary. 
Note that the delta function squared means that not only the energy
density but also the total energy emitted diverge. 
Such a divergent flux suggests that its backreaction to the spacetime
and boundary is not ignorable. See Fig.~\ref{fg:VEV} for 3D plots of the
energy density and momentum density with cutoff. 

We have to pay attention also to the second term on the right-hand side of Eq.~\eqref{Tab_03}. Everywhere except the null lines emanating from the appearance
point, this divergent summation has the same form as Eq.~\eqref{Tab_02}.
Hence, by the same renormalization procedure, we obtain a finite result for $t>0$ which is the same as Eq.~\eqref{eq:simple_Casimir} for $t<0$. On the other hand, on the null lines, the summation is different from the previous one. Such a different divergence may yield a non-renormalizable ultraviolet divergence.
Indeed, we present another derivation of Eq.~\eqref{Tab_03} by the
Green-function method in Appendix~\ref{sec:green} and evaluate 
``a renormalized energy-momentum tensor'' by the point-splitting method.
The result gives Eq.~(\ref{eq:simple_Casimir}) for $t<0$ but 
the following for $t>0$:
\be
    \langle 0_g | {\bm T}_{\pm\pm} | 0_g \rangle_{{\rm ren}, t>0}=
	\frac{  1 }{\pi}
	\sum_{ \substack{ m=1 \\ m:{\rm odd} } }^\infty
	\frac{ 1 }{m}
	\sum_{\ell=-\infty}^\infty \delta( z_\pm - \ell L )^2
     +
    \begin{cases}
    \displaystyle
     - \frac{\pi}{24L^2}
     + \lim_{\Delta z_\pm \to 0} \frac{1}{8\pi\Delta z_\pm^2}
     & (z_\pm = \ell L, \; \ell \in {\bf Z}) \\
     \displaystyle
     - \frac{\pi}{48L^2}  & ({\rm otherwise}) \\
    \end{cases} .
\label{Tab_03_ren}
\ee
Thus, in the coincidence limit $\Delta z_\pm \to 0$, an ultraviolet divergence like $(\Delta z_\pm)^{-2}$ remains on the null lines.
The divergence like $(\Delta z_{\pm})^{-2}$ suggests that the contribution to the total energy from this term also diverges.

\section{Sudden disappearance of a Dirichlet boundary}
\label{sec:disapp}

\begin{figure}[bth]
\begin{center}
\begin{minipage}[c]{0.8\textwidth}
\linespread{1}
\begin{center}
\includegraphics[height=6cm]{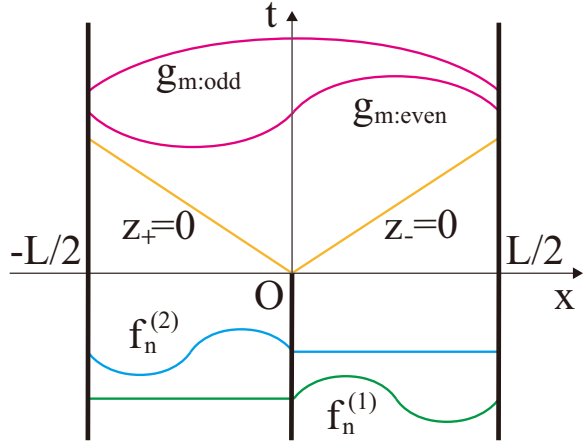}
\caption{The sudden disappearance of a Dirichlet boundary in a 1D
 cavity. The scalar field obeys the Dirichlet boundary conditions at the
 both ends $(x=\pm \frac{L}{2})$ for $-\infty<t<\infty$ and at the
 center $(x=0)$ for $t < 0$. The null lines $z_\pm := t \pm x = 0 \; (t>0)$ and the spatial configurations of mode functions $g_m$ and $f_n^{(\gamma)} \; (m,n \in {\bf N}, \gamma \in \{1,2 \})$ are schematically depicted.
}
\label{fg:bc2}
\end{center}
\end{minipage}
\end{center}
\end{figure}

In this section, we consider the sudden disappearance of the Dirichlet boundary (see Fig.~\ref{fg:bc2}). Since the situation is a kind of time reversal of that in the previous section, most parts of calculation can be reused in this section. 

\subsection{Classical behaviors and quantization of the massless scalar field}

In addition to the Dirichlet boundary conditions at the both ends~\eqref{BC_end}, the scalar field obeys the Dirichlet boundary condition at the center before $t=0$,
 \be
	\phi (t,0)=0, \;\;\;	 t<0.
\label{BC_DisApp}
\ee
Then, the positive-energy mode functions in the asymptotic regions $t \to -\infty$ and $t \to \infty$ are given by Eqs.~\eqref{f} and \eqref{g}, respectively.

We expand $f_n^{(\gamma)}$ by $g_m$, 
\be
	f_n^{(\gamma)}
	=
	\sum_{m=1}^\infty ( \alpha_{nm}^{(\gamma)} g_m + \beta_{nm}^{(\gamma)} g_m^\ast ).
\label{f_g}
\ee
Here, the expansion coefficients are given by
\be
	\alpha_{nm}^{(\gamma)}
	=
	\langle g_m, f_n^{(\gamma)} \rangle
	=
	\rho_{mn}^{(\gamma)\ast},
\;\;\;
	\beta_{nm}^{(\gamma)}
	=
	- \langle g_m^\ast, f_n^{(\gamma)} \rangle
	=
	-\sigma_{mn}^{(\gamma)},
\label{alpha_beta}
\ee
where $\rho_{mn}^{(\gamma)}$ and $\sigma_{mn}^{(\gamma)}$ are given by Eq.~\eqref{rho_sigma}.

The quantization of the scalar field is again implemented by
Eqs.~\eqref{phi_g}--\eqref{CR_a}. Substituting Eq.~\eqref{f_g} into
Eq.~\eqref{phi_f} and comparing it with \eqref{phi_g}, we obtain
\be
	{\bm b}_m
	=
	\sum_{\gamma=1}^2 \sum_{n=1}^\infty
	(
		\alpha_{nm}^{(\gamma)} {\bm a}^{(\gamma)}_n
		+
		\beta_{nm}^{(\gamma) \ast } {\bm a}^{(\gamma)\dagger}_n
	).
\label{bBYa}
\ee
Substituting Eq.~\eqref{bBYa} into Eq.~\eqref{CR_b}, and using Eq.~\eqref{CR_a}, one finds that the following conditions must hold for the two quantizations, Eqs.~\eqref{phi_g} and \eqref{phi_f}, to be consistent.  
\begin{align}
	\sum_{\gamma=1}^2 \sum_{n=1}^\infty
	(
		\alpha_{nm}^{(\gamma)} \alpha_{nm'}^{(\gamma)\ast}
		-
		\beta_{nm}^{(\gamma)\ast} \beta_{nm'}^{(\gamma)}
	)
	&=
	\delta_{mm'},
\label{Consis3}
\\
	\sum_{\gamma=1}^2 \sum_{n=1}^\infty
	(
		\alpha_{nm}^{(\gamma)} \beta_{nm'}^{(\gamma)\ast}
		-
		\beta_{nm}^{(\gamma)\ast} \alpha_{nm'}^{(\gamma)}
	)
	&=
	0.
\label{Consis4}
\end{align}
It is shown in Appendix~\ref{sec:con2} that these unitarity relations indeed hold for the Bogoliubov coefficients given by Eq.~\eqref{alpha_beta}. 

Since we are interested in the particle creation due to the disappearance of boundary, we assume that the quantum field is in the vacuum state $| 0_f \rangle$ in which any particle corresponding to $f_n^{(\gamma)}$ does not exist. Such a vacuum is characterized by
\be
	{\bm a}_n^{(\gamma)} |0_f \rangle = 0,
\;\;\;
	\langle 0_f | 0_f \rangle =1,
\;\;\;
	\forall \gamma \in \{ 1,2 \},
\;\;\;
	\forall n \in {\bf N}.
\label{vac2}
\ee

\subsection{Spectrum and energy-momentum density}

The vacuum $| 0_f \rangle$ contains no particle corresponding to $f_n^{(\gamma)}$ but can contain particles corresponding to $g_m$. This is examined by calculating the vacuum expectation value of particle-number operator,
\begin{align}
	\langle 0_f |  {\bm b}_m^\dagger {\bm b}_m | 0_f \rangle
	=
	\sum_{\gamma=1}^2 \sum_{n=1}^\infty | \beta_{nm}^{(\gamma)} |^2
	=
	\begin{cases}
	\displaystyle \frac{2}{m\pi^2} \sum_{n=1}^\infty \frac{n}{(n+m/2)^2} & (m:\mbox{odd}) \\
	\displaystyle  0 & (m:\mbox{even})
	\end{cases}.
\label{NumDen4}
\end{align}
This is logarithmically divergent for odd $m$. Therefore, the total
number of created particles, i.e.\ the summation over $m \in {\bf N}$ of
Eq.~\eqref{NumDen4}, also diverges. This implies again that the Fock
space representation associated with ${\bm a}_n^{(\gamma)}$ is unitarily
inequivalent to that associated with ${\bm b}_{m}$~\cite{Wald:1995yp,Rodriguez-Vazquez:2014hka}.

Substituting Eq.~\eqref{phi_f} into Eq.~\eqref{tpm}, and using
Eq.~\eqref{f}, we obtain the vacuum expectation value of the
energy-momentum tensor before the disappearance of the Dirichlet boundary,
\begin{align}
	\langle 0_f | {\bm T}_{\pm \pm} |0_f \rangle_{t<0}
	=
	\sum_{\gamma=1}^2 \sum_{n=1}^\infty | \pd_{\pm} f_n^{(\gamma)} |^2
	=
	\frac{\pi}{L^2} \sum_{n=1}^\infty n.
\label{Tab_12}
\end{align}
We can renormalize this by standard procedures again to obtain a finite
result,
\begin{equation}
\langle 0_f | {\bm T}_{\pm \pm} |0_f \rangle_{{\rm ren}, t<0} = - \frac{\pi}{12L^2}. 
\label{eq:simple_Casimir_2}
\end{equation}

What we are most interested in is the same quantity after the disappearance of the Dirichlet boundary. Substituting Eq.~\eqref{phi_g} into Eq.~\eqref{tpm}, and then using Eq.~\eqref{bBYa}, such a quantity is obtained as
\begin{align}
	 &\langle 0_f |{\bm T}_{\pm\pm}| 0_f \rangle_{t>0}
\nn
\\
	 &=
	 \sum_{\gamma=1}^2
	 \sum_{n=1}^\infty
	 \sum_{ \substack{ m=1 \\ m:{\rm odd} } }^\infty
	 \sum_{ \substack{ m'=1 \\ m':{\rm odd} } }^\infty
	 [
		(
			\alpha_{nm}^{(\gamma)} \beta_{nm'}^{(\gamma)}
			+
			\alpha_{nm'}^{(\gamma)} \beta_{nm}^{(\gamma)}
		) {\rm Re} (\pd_\pm g_m \pd_\pm g_{m'})
		+
		(
			\alpha_{nm}^{(\gamma)} \alpha_{nm'}^{(\gamma)}
			+
			\beta_{nm'}^{(\gamma)} \beta_{nm}^{(\gamma)}
		) {\rm Re} (\pd_\pm g_m \pd_\pm g^\ast_{m'})
	 ]
\nn
\\
	&+
	\sum_{\gamma=1}^2
	\sum_{n=1}^\infty
	\sum_{ \substack{ m=2 \\ m:{\rm even} } }^\infty
	\sum_{ \substack{ m'=2 \\ m': {\rm even} } }^\infty
	\alpha_{nm}^{(\gamma)} \alpha_{nm'}^{(\gamma)}
	{\rm Re} (\pd_\pm g_m \pd_\pm g^\ast_{m'}).
\label{TAB_11}
\end{align}
To derive Eq.~\eqref{TAB_11}, we symmetrize the dummy indices $m$ and $m'$. In addition, we use the facts that $ \beta_{nm}^{(\gamma)} $ vanishes for even $m$, and implicitly use a few properties of Bogoliubov coefficients~\eqref{alpha_beta} such as the $\gamma$-dependence. 

Using the explicit form of Bogoliubov coefficients and mode functions, Eqs.~\eqref{alpha_beta}, \eqref{rho_sigma}, and \eqref{g}, equation \eqref{TAB_11} is written as
\begin{align}
	 \langle 0_f |{\bm T}_{\pm\pm}| 0_f \rangle_{t>0}
	=
	 \frac{8}{\pi L^2}
	 \sum_{n=1}^\infty
	 \left(
	  4n^3
	 [
		\sum_{ \substack{ m=1 \\ m: {\rm odd} } }^\infty
		\frac{ \cos ( \frac{m\pi}{L} z_\pm ) }{ m^2-(2n)^2 }
	]^2
	+
	n
	 [
		\sum_{ \substack{ m=1 \\ m: {\rm odd} } }^\infty
		\frac{ m \sin ( \frac{m\pi}{L} z_\pm ) }{ m^2-(2n)^2 }
	]^2
	\right)
	+ \frac{\pi}{2L^2}
	\sum_{n=1}^\infty n.
\label{TAB_12}
\end{align}
This is an even function of $z_\pm$ with period $L$, as it is invariant under reflection $z_\pm \to - z_\pm$ and translation $z_\pm \to z_\pm + L$. Therefore, it is sufficient to calculate it in $0 \leq z_\pm < L$, and then generalize the obtained expression to one valid in the entire domain appropriately.

The first summation over odd $m$ in Eq.~\eqref{TAB_12} can be calculated to give
\begin{align}
	 \langle 0_f |{\bm T}_{\pm\pm}| 0_f \rangle_{t>0}
	=
	 \frac{8}{\pi L^2}
	 \sum_{n=1}^\infty
	 n \left(
	 \frac{\pi^2}{16} \sin^2 ( \frac{2n\pi}{L} z_\pm )
	+
	 [
		\sum_{ \substack{ m=1 \\ m: {\rm odd} } }^\infty
		\frac{ m \sin ( \frac{m\pi}{L} z_\pm ) }{ m^2-(2n)^2 }
	]^2
	\right)
	+ \frac{\pi}{2L^2}
	\sum_{n=1}^\infty n,
\label{TAB_13}
\end{align}
which is valid in $0 \leq z_\pm < L$. Here, we have used the following formula~\cite[p.\ 733]{maru},
\begin{align}
	\sum_{k=0}^\infty
	\frac{ \cos [(2k+1)y] }{(2k+1)^2 -a^2} 
	=
	\frac{\pi}{4a} \sin[ \frac{a}{2}(\pi - 2y) ] \sec (\frac{a\pi}{2}),
\;\;\;
	( 0 \leq y \leq \pi ).
\label{maru_typo1}
\end{align}
It is noted here that there are typos in Ref.~\cite[p.\ 733]{maru} about formulas~\eqref{maru_typo1} and \eqref{maru_typo2} (see below).

For $z_\pm=0$, from Eq.~\eqref{TAB_13}, we have
\be
	\langle 0_f |{\bm T}_{\pm\pm}| 0_f \rangle_{t>0}
	=
	\frac{\pi}{2L^2}
	\sum_{n=1}^\infty n,
\;\;\;
	(z_\pm = 0).
\label{TAB_14}
\ee

For $ 0 < z_\pm < L$, we find that the summation over odd $m$ in Eq.~\eqref{TAB_13} can be calculated to give
\be
	\langle 0_f |{\bm T}_{\pm\pm}| 0_f \rangle_{t>0}
	=
	\frac{\pi}{L^2}
	\sum_{n=1}^\infty n,
\;\;\;
	(0 < z_\pm < L),
\label{TAB_15}
\ee
using the following formula~\cite[p.\ 733]{maru},
\begin{align}
	\sum_{k=0}^\infty
	\frac{ (2k+1) \sin [(2k+1)y] }{(2k+1)^2 -a^2} 
	&=
	\frac{\pi}{4} \cos[ \frac{a}{2}(\pi - 2y) ] \sec (\frac{a\pi}{2}),
\;\;\;
	( 0 < y < \pi ).
\label{maru_typo2}
\end{align}

Extending the domain of Eqs.~\eqref{TAB_14} and \eqref{TAB_15} to the entire domain periodically, 
we obtain
\be
	\langle 0_f | {\bm T}_{\pm \pm} |0_f \rangle_{t>0}
	=
	\begin{cases}
	\displaystyle \frac{\pi}{2L^2} \sum_{n=1}^\infty n & (z_\pm = \ell L, \; \ell \in {\bf Z} )\\
	\displaystyle \frac{\pi}{L^2} \sum_{n=1}^\infty n & ({\rm otherwise})\\
	\end{cases} .
\label{Tab_13}
\ee

Although we have no term proportional to the delta function squared in contrast to the appearance case, we have to pay attention again to the diverging summations in Eq.~\eqref{Tab_13}. On the null lines, the energy-momentum tensor \eqref{Tab_13} takes the different form from Eq.~\eqref{Tab_12}. Indeed, the renormalized energy-momentum tensor is given by Eq.~(\ref{eq:simple_Casimir_2}) for $t<0$ but by the following for $t>0$: 
\be
    \langle 0_f | {\bm T}_{\pm\pm} | 0_f \rangle_{{\rm ren}, t>0}=
    \begin{cases}
    \displaystyle
	 - \frac{\pi}{24L^2}
     + \lim_{\Delta z_\pm \to 0} \frac{1}{8\pi\Delta z_\pm^2}
     & (z_\pm = \ell L, \; \ell \in {\bf Z}) \\
     \displaystyle
     - \frac{\pi}{12L^2} 
     & ({\rm otherwise}) \\
    \end{cases},
\label{Tab_14}
\ee
which diverges on the null lines in the coincidence limit $\Delta z_\pm \to 0$. See Appendix~\ref{sec:green} for the derivation of this result using the Green functions. 

The above ultraviolet divergence on the null lines seems to play a physically
significant role as follows. After the Dirichlet wall disappears and the cavity size becomes 
$L$ for $t>0$, the ambient Casimir energy density remains the same as the energy density with the cavity size $L/2$ for $t<0$. This means that the amount of energy for $t>0$ would be lower than that of the ground state with the cavity size $L$ if the divergence on the null lines was not taken into account. Thus, it is expected that this divergent flux would compensate for the shortage of energy in the cavity. This expectation holds if the total energy radiated on the null lines diverges due to the term proportional to $(\Delta
z_{\pm})^{-2}$.

\section{Instantaneous limit of smooth appearance and disappearance}
\label{sec:smooth}

As we have seen, the sudden appearance and disappearance of the
Dirichlet wall will cause the different behaviors of ultraviolet
divergence for the energy-momentum tensor.
For understanding their origins, it would be helpful to compare results
in smooth appearance and disappearance models.

In Ref.~\cite{Brown:2015yma}, the authors investigated the vacuum
excitation by the smooth appearance of a both-sided Dirichlet wall in
$1+1$ dimensional Minkowski spacetime and its instantaneous limit. 
Therefore, following their formulation, we will see that the divergent
behaviors are quite similar to those observed in Secs.~\ref{sec:app} and
\ref{sec:disapp} after taking a certain limit of smooth appearance and
disappearance models.

Let us briefly review the formulation and result of
Ref.~\cite{Brown:2015yma} in Sec.~\ref{BL}. Then, in Sec.~\ref{SA}, we
will show  that the divergent behavior such as the delta function
squared and the ultraviolet divergence on the null lines appear by
taking the instantaneous limit of its model. Also, in Sec.~\ref{SD}, we generalize the formulation 
to the disappearance case and consider its instantaneous limit.

In this section, for simplicity, we leave the existence of the cavity boundaries out of consideration because we are only interested in the ultraviolet divergent behavior of the energy-momentum tensor independent of cavity size $L$.
In addition, we focus on the even-parity modes of the scalar field
because the odd-parity modes are irrelevant to existence or absence of
the Dirichlet wall at $x=0$.

\subsection{Smooth-appearance model}
\label{BL}

In Ref.~\cite{Brown:2015yma}, for analyzing a smooth appearance of the
Dirichlet wall, the authors introduce the $\delta$-function potential with a smooth time-dependent coefficient into the Klein-Gordon equation of motion,
\begin{equation}
 \left[\partial^2_t - \partial^2_x + 
  \frac{2\cot(\theta(t))}{ {\cal L} }\delta(x)\right]\phi = 0 ,
 \label{eq:KG_pot}
\end{equation}
where ${\cal L}$ is a positive constant. They assume that function $\theta$ in the coefficient is given by
\begin{equation}
 \theta(t) = \arctan\left(\frac{1+e^{-\lambda t}}{\lambda {\cal L} }\right),
\label{theta1}
\end{equation} 
where $\lambda$ is a positive constant. This choice of $\theta$ corresponds to the following time-dependent boundary condition at $x=0$,
\be
		\pd_x \phi(t,x)|_{x=0+} = \frac{ \lambda  }{ 1+e^{- \lambda t} } \phi(t,x) |_{x=0+},
\label{SA_BC}
\ee
where we have used $\phi(t,x)$ is an even function with respect to $x=0$. Note that boundary condition~\eqref{SA_BC} is obtained by
integrating Eq.~\eqref{eq:KG_pot} over an infinitesimal interval across $x=0$ and substituting Eq.~\eqref{theta1} into it. One can see that boundary condition~\eqref{SA_BC} reduces to the Neumann one in the asymptotically past $t \to -\infty$ and to the Dirichlet one in the asymptotically future $t \to +\infty$ as long as $\lambda$ is sufficiently large. $\lambda^{-1}$ represents the time scale of appearance process, and therefore the limit of $\lambda \to \infty$ corresponds to the instantaneous limit of smooth appearance of a Dirichlet wall. 

A set of positive-energy mode function $ \{  U_k \} \; ( k>0) $ is written in the following form
\be
	U_k (z_-,z_+) = \frac{1}{\sqrt{8\pi k}} \left[e^{-ik z_+} + E_k( z_- )\right].
\label{eq:Uk}
\ee
Substituting Eq.~\eqref{eq:Uk} into Eq.~\eqref{SA_BC} and imposing the regularity at $t \to \infty$, one obtains
\be
	E_k ( z_- )   = \frac{e^{-ik z_- }}{1+e^{\lambda z_-}}
  \left(1- \frac{\lambda + ik}{\lambda - ik} e^{\lambda z_-}\right).
  \label{eq:Ek}
\ee

When (the even sector of) field operator ${\bm \phi}$ is expanded as 
\begin{equation}
 {\bm \phi}
  =
  \int^\infty_0 dk 
  ({\bm a}_k U_k + {\bm a}_k^\dagger U_k^\ast) ,
\end{equation}
a non-vanishing component of the renormalized energy-momentum tensor is given by 
\begin{equation}
 \langle 0 | {\bm T}_{--} | 0 \rangle_{\rm ren} = \int_\mu^\infty \frac{dk}{8\pi k}
  \left(|E'_k(z_-)|^2 - k^2\right) ,
 \label{BL-vev}
\end{equation}
where $|0 \rangle$ denotes an ordinary Minkowski vacuum, which is annihilated by all right-propagating and left-propagating positive-energy modes, and $\mu$ is an infrared cutoff introduced by hand. The second term ($-k^2$) in the integrand of Eq.~\eqref{BL-vev} is the subtraction term for the vacuum expectation value in the Minkowski spacetime.

Substituting Eq.~\eqref{eq:Ek} into Eq.~\eqref{BL-vev}, one obtains
\be
	\langle 0 | {\bm T}_{--} | 0 \rangle_{\rm ren}
	=
	\frac{ \lambda^2 \ln [ 1+(\lambda/\mu)^2 ] }{ 64\pi \cosh^4 ( \lambda z_-/2 ) }.
\ee
It is clear that in the limit of $\lambda \to \infty$ this quantity diverges on the null line $z_-=0$ and vanishes on $z_- \neq 0$. In order to estimate the strength of divergence on $z_-=0$, the above expression is rewritten as
\be
	\langle 0 | {\bm T}_{--} | 0 \rangle_{\rm ren}
	=
	\frac{ \lambda \ln [ 1+(\lambda/\mu)^2 ] }{ 24\pi  }
	\delta_\lambda ( z_- ),
\;\;\;
	\delta_\lambda (z_-) := \frac{ 3\lambda }{ 8 \cosh^4( \lambda z_-/2 ) }.
\ee
Taking into account that $ \lim_{\lambda \to \infty}  \delta_\lambda (z_-) = \delta (z_-) $ (see Eq.~\eqref{acosh^k} below), Ref.~\cite{Brown:2015yma} concluded that the divergence on the null line $z_-=0$ of $ \langle 0 | {\bm T}_{--} | 0 \rangle_{\rm ren} $ in the instantaneous limit is too strong to have a distributional limit.

\subsection{Smooth appearance and instantaneous limit}
\label{SA}

Now, we will see that the divergent behavior similar to that observed in Sec.~\ref{sec:app} can be obtained by taking an instantaneous limit of the above smooth appearance model.

Note that Ref.~\cite{Brown:2015yma} takes the instantaneous limit
$\lambda \to \infty$ after computing the momentum integration in
Eq.~\eqref{BL-vev}. Instead, we take the instantaneous limit before the
momentum integration. Using Eq.~\eqref{eq:Ek}, we compute the integrand
in Eq.~\eqref{BL-vev} in an instantaneous regime $ k /\lambda \ll 1$%
\footnote{
As we mentioned, in the current model, we should take sufficiently large
$\lambda$ to realize the Dirichlet boundary condition at the asymptotic future.
In fact, the asymptotic form becomes
$E_k(z_-) \sim - e^{-ikz_-} (\lambda - i k)/(\lambda + i k)$ as
$t\to\infty$ rather than $E_k(z_-) \sim - e^{-ikz_-}$.
Therefore, $k/\lambda \ll 1$ must be kept even if momentum $k$ becomes large.
}
as
\begin{equation}
 |E'_k(z_-)|^2 - k^2 = \frac{\lambda^2}{4\cosh^4(\lambda z_-/2)} 
  - \frac{ k^2 }{ 4\cosh^4( \lambda z_- /2) } 
  \big[ 1+ \mathcal{O}(\frac{ k^2 }{ \lambda^2 } ) \big],
\end{equation}
where $ \mathcal{O}(\frac{ k^2 }{ \lambda^2 } ) $-term has no dependence on $z_-$. 
Taking the limit $\lambda \to \infty$ of the above, we have
\begin{equation}
 |E'_k(z_-)|^2 -k^2 = 4 \delta (z_-)^2 -  
     \begin{cases}
    \displaystyle
	 \frac{k^2}{4}
     & (z_- = 0) \\
     \displaystyle
     0
     & ({\rm otherwise}) \\
    \end{cases},
\;\;\;
	( \lambda \to \infty),
\label{BL-integrand}
\end{equation}
where we have used the following mathematical relations
\be
&&
	\lim_{a \to \infty} \frac{  a }{  \cosh^k ( a y )} = \frac{ 2^k[ (k-2)!! ]^2 }{ (2k-2)!! } \delta (y),
\;\;\;
	k=2,4,6,\cdots,
\label{acosh^k}
\\
&&
	\lim_{a \to \infty} \frac{ 1 }{ \cosh^k (ay) }
	=
	\begin{cases}
    \displaystyle
	 1
     & (y = 0) \\
     \displaystyle
     0
     & ( y \neq 0 ) \\
    \end{cases},
 \;\;\;
	k=1,2,3,\cdots.
\ee
Substituting Eq.~\eqref{BL-integrand} into Eq.~\eqref{BL-vev}, we have
\be
	\langle 0 | {\bm T}_{--} | 0 \rangle_{\rm ren} 
	=
	\frac{ \delta(z_-)^2 }{ 2\pi } \int_\mu^\infty \frac{dk}{k} 
	-
	\begin{cases}
    \displaystyle
	 \frac{1}{32\pi} \int_\mu^\infty dk k
     & (z_- = 0) \\
     \displaystyle
     0
     & ({\rm otherwise}) \\
    \end{cases},
\;\;\;
	( \lambda \to \infty).
\ee

Thus, we have obtained the delta function squared multiplied by a logarithmically divergent factor and the ultraviolet divergence that exists only on the null line $z_-=0$. Note that divergent integral $\int_\mu^\infty dk k$ turns out to be proportional to $-(\Delta z_-)^{-2}$ if one adopts the point-splitting regularization. 

\subsection{Smooth disappearance and instantaneous limit}
\label{SD}

Here, let us generalize the argument of smooth appearance of the Dirichlet wall in Ref.~\cite{Brown:2015yma} to the smooth disappearance of Dirichlet wall. Then, we will consider its instantaneous limit.

In order to model the disappearance of Dirichlet wall in the formulation, we consider the time reversal $t \to -t$ of smoothing function \eqref{theta1} as
\be
	\theta(t)
	=
	\arctan \left( \frac{ 1+e^{\lambda t} }{ \lambda {\cal L} } \right),
\;\;\;
	(\lambda >0),
\label{theta2}
\ee
which corresponds to the following time-dependent boundary condition at the center,
\be
	\pd_x \phi(t,x)|_{x=0+} = \frac{ \lambda  }{ 1+e^{\lambda t} } \phi(t,x) |_{x=0+}.
\label{SD_BC}
\ee

Substituting the ansatz of mode function~\eqref{eq:Uk} into Eq.~\eqref{SD_BC} and imposing the regularity at $ t \to -\infty$, we obtain
\be
	E_k (z_-)
	=
	-e^{ -ik z_- }
	-  \frac{ 2ik }{ \lambda -i k } (1+e^{ \lambda z_- }) 
	{}_2 F_1 \left( 1,1- i \frac{k}{\lambda}, 2- i \frac{k}{\lambda} ;  -e^{ \lambda z_- } \right) e^{-ik z_-},
\label{eq:Ek2}
\ee
where ${}_2 F_1$ is the hypergeometric function. 

Using Eq.~\eqref{eq:Ek2}, we compute the integrand in Eq.~\eqref{BL-vev} in the instantaneous regime $ k /\lambda \ll 1$,
\be
	| E_k' (z_-) |^2 -k^2
	=
	-4
	e^{-\lambda z_-} \ln ( 1+e^{\lambda z_-} ) \cdot
	[ 1-e^{-\lambda z_-} \ln ( 1+e^{\lambda z_-} ) ] k^2 + {\cal O} (\frac{ k^3 }{ \lambda^3 } ) .
\ee
If we take the instantaneous limit $\lambda \to \infty$ of the above, we obtain
\be
	| E_k' (z_-) |^2 -k^2
	=
	\begin{cases}
		- 4 (1-\ln 2) \ln 2 \cdot k^2& (z_- = 0) \\
		0 & (\mbox{otherwise}) \\
	\end{cases},
\;\;\;
	( \lambda \to \infty ),
\label{BL-integrand2}
\ee
using the following,
\be
	\lim_{ a \to \infty } e^{-ay} \ln ( 1+e^{ay} )
	=
	\begin{cases}
		1 & (y<0) \\
		\ln 2 & (y=0) \\
		0 & (y>0) \\
	\end{cases}.
\ee
Substituting Eq.~\eqref{BL-integrand2} into Eq.~\eqref{BL-vev}, we obtain
\be
	\langle 0 | {\bm T}_{--} | 0 \rangle_{\rm ren}
	=
	\begin{cases}
		\displaystyle -\frac{  (1-\ln 2) \ln 2}{ 2\pi }\int_\mu^\infty dk k & (z_-=0) \\
		0 & (\mbox{otherwise}) \\
	\end{cases},
\;\;\;
	( \lambda \to \infty ).
\ee

From the above expression, we observe that the term of delta function squared is absent, and only the ultraviolet divergence that exists only on the null line $z_-=0$ appears. Thus, we have obtained the divergent energy-momentum tensor of which main features are the same as those in Sec.~\ref{sec:disapp}.

\section{Conclusion}
\label{sec:conc}

We have investigated the vacuum excitation of a massless
Klein-Gordon scalar field due to the sudden appearance and disappearance
of a both-sided Dirichlet wall in a 1D cavity. 

For the sudden appearance of the Dirichlet wall, we found that the
vacuum is highly excited to result in the infinitely strong flux given
by Eq.~\eqref{Tab_03_ren}. This result suggests that the backreaction to the background spacetime and boundary cannot be ignored. In other words, the background spacetime is forced to be dynamical and/or the instantaneous insertion of the Dirichlet wall itself is prohibited by the quantum field. We note that the result is quite similar to those in
the investigation of the topology change~\cite{Anderson:1986ww,Manogue}
and the strong curvature singularity~\cite{Ishibashi:2002ac}, although the boundary condition in the present work is different from those in the papers.

Also for the sudden disappearance of the Dirichlet wall, we
found that the vacuum is highly excited to result in the infinitely strong flux given by Eq.~\eqref{Tab_14}. In contrast to the appearance case, the renormalized energy-momentum tensor does not contain the term proportional to the delta function squared, although it contains the diverging term proportional to $(\Delta z_{\pm})^{-2}$.
The infinite flux is what we expect from the viewpoint of the number of created particles as mentioned in Introduction, while the lack of the delta function squared is not. 

Let us mention the divergence of the renormalized energy-momentum tensor appearing both in the sudden appearance and disappearance cases. We have seen that the standard procedure of the point-splitting regularization gives a finite value of the renormalized energy-momentum tensor at the spacetime points not on the null lines which emanate from the transition point, while it is divergent on the null lines. We have interpreted this result as the diverging flux on the null lines for $t>0$. While there seems no ambiguity in this straightforward interpretation, it is more convincing if such a peculiar divergence on null lines appears as the result of an instantaneous limit of finite-time appearance and disappearance of the Dirichlet wall. Therefore, using the formulation in Ref.~\cite{Brown:2015yma}, which estimates the particle creation by a smoothly appearing Dirichlet wall, we have shown in Sec.~\ref{sec:smooth} that the $(\Delta z_\pm)^{-2}$-type divergence appears on the null lines after taking an instantaneous limit for the appearance and disappearance cases, although the discussion is restricted only to the infinite cavity limit $(L \to \infty)$.

The discrepancy between the appearance and disappearance cases seems to
stem from the different behaviors of two sets of mode functions,
$\{f_n^{(\gamma)}\}$ and $\{ g_m \}$, which define distinct vacua $|0_f \rangle$ and $ |0_g \rangle $, respectively.
First, let us see the behavior of $f_n^{(\gamma)}$. While $f_n^{(\gamma)}$ is given by
Eq.~\eqref{f} for $t<0$, it is expressed as Eq.~\eqref{f_g} for
$t>0$. Here, the point is that $ f_n^{(\gamma)} $ given by
Eqs.~\eqref{f} and \eqref{f_g} coincide in the limit of $t \to 0$, which
implies that $f_n^{(\gamma)}$ is continuous at $t=0$. Next, let us see
the behavior of $ g_m $. While $g_m$ is given by Eq.~\eqref{g} for
$t<0$, $g_m$ is expressed as Eq.~\eqref{gBYf} for $t>0$. In this case,
$g_m$ given by Eqs.~\eqref{g} and \eqref{gBYf} do not coincide in the
limit of $t \to 0$ at every point of $[-L/2, L/2]$. Namely, when $m$ is
odd, while $\lim_{t \to -0} g_m(t,0) \neq 0$ from Eq.~\eqref{g},
$\lim_{t \to +0} g_m(t,0) = 0$ from Eq.~\eqref{gBYf} (note that the
right-hand side of Eq.~\eqref{gBYf} consists only of sine functions),
which implies the discontinuity of $g_m \; (m \in {\rm odd})$ at
$t=0$. We conjecture that the existence of such a discontinuity of mode
functions is the origin of the square of the delta function in the appearance case.

Given the results in this paper, we have many things to examine. In
particular, it is important to prove (or disprove) that
the present result
is not an artifact of simplification and idealization adopted in our analysis (i.e., equal lengths of left and right regions, low dimensionality, scalar field, and so on). The generalizations of this work in this direction will be indispensable to understand how much the semiclassical effects play crucial roles in the gravitational phenomena such as the spacetime connection and disconnection. 

\subsection*{Acknowledgments}
The authors would like to thank an anonymous referee for suggesting us to compare the result in this paper with that of Ref.~\cite{Brown:2015yma}, which deepened our understanding about the current topic. UM would like to thank H.\ Maeda, A.\ Ishibashi, and H.\ Okamoto for useful discussions. This work was supported by JSPS KAKENHI Grant 
Numbers 26400282 (TH) and 15K05086 (UM).

\appendix
\section{Proof of unitarity relations~\eqref{Consis1} and \eqref{Consis2}}
\label{sec:con1}

Substituting Eq.~\eqref{rho_sigma} into the left-hand side of Eq.~\eqref{Consis1}, we obtain
\be
	\sum_{m=1}^\infty
	(
		\rho_{mn}^{(\gamma)} \rho_{mn'}^{(\gamma') \ast}
		-
		\sigma_{mn}^{(\gamma)\ast} \sigma_{mn'}^{(\gamma')}
	)
	=
	\frac{16 \sqrt{nn'} (n+n')}{ \pi^2 }
	\sum_{ \substack{ m=1 \\ m:{\rm odd} }  }^\infty
	\frac{1}{[ m^2-(2n)^2 ] [ m^2-(2n')^2 ]}
	+
	\frac12 (-1)^{\gamma+\gamma'} \delta_{nn'}.
\nn
\\
\label{sum01}
\ee
The summation over odd $m$ in Eq.~\eqref{sum01} can be evaluated to give
\be
	\sum_{ \substack{ m=1 \\ m:{\rm odd} }  }^\infty
	\frac{1}{[ m^2-(2n)^2 ] [ m^2-(2n')^2 ]}
	=
	\frac{\pi^2}{16(2n)^2} \delta_{nn'},
\label{sum02}
\ee
using the following formulas~\cite[pp.\ 688--689]{maru}
\begin{align}
	\sum_{k=0}^\infty \frac{1}{(2k+1)^2-a^2}
	&=
	\frac{\pi}{4a} \tan( \frac{a \pi }{2} ),
\label{form688-1}
\\
	\sum_{k=0}^\infty
	\frac{1}{[(2k+1)^2-a^2]^2}
	&=
	-\frac{\pi}{8a^3}\tan (\frac{a\pi}{2}) + \frac{\pi^2}{16a^2}{\rm sec}^2( \frac{a\pi}{2} ).
\label{form689-1}
\end{align}
Substituting Eq.~\eqref{sum02} into Eq.~\eqref{sum01}, we see Eq.~\eqref{Consis1} to hold.

Substituting Eq.~\eqref{rho_sigma} into the left-hand side of Eq.~\eqref{Consis2}, we obtain
\be
	\sum_{m=1}^\infty
	(
		\rho_{mn}^{(\gamma)} \sigma_{mn'}^{(\gamma') \ast}
		-
		\sigma_{mn}^{(\gamma)\ast} \rho_{mn'}^{(\gamma')}
	)
	=
	-\frac{16\sqrt{nn'}(n-n') }{\pi^2}
	\sum_{ \substack{ m=1 \\ m:{\rm odd} }  }^\infty
	\frac{1}{[ m^2-(2n)^2 ] [ m^2-(2n')^2 ]}.
\label{sum03}
\ee
Substituting Eq.~\eqref{sum02} into Eq.~\eqref{sum03}, we see Eq.~\eqref{Consis2} to hold.

\section{Proof of unitarity relations~\eqref{Consis3} and \eqref{Consis4}}
\label{sec:con2}

We define
\be
	I^{(\gamma)}_{ mm^\prime }
	:=
	\sum_{n=1}^\infty
	( \alpha_{nm}^{(\gamma)} \alpha_{nm^\prime}^{(\gamma) \ast} - \beta_{nm}^{(\gamma)\ast} \beta_{nm^\prime}^{(\gamma)} ),
\;\;\;
	J^{(\gamma)}_{ mm^\prime }
	:=
	\sum_{n=1}^\infty
	( \alpha_{nm}^{(\gamma)} \beta_{nm^\prime}^{(\gamma) \ast} - \beta_{nm}^{(\gamma)\ast} \alpha_{nm^\prime}^{(\gamma)} ).
\label{IJ}
\ee
Then, the unitarity relations \eqref{Consis3} and \eqref{Consis4} are rewritten as
\be
	\sum_{\gamma=1}^2  I^{(\gamma)}_{ mm^\prime } = \delta_{mm'},
\;\;\;
	\sum_{\gamma=1}^2 J^{(\gamma)}_{ mm^\prime } = 0.
\label{uni3}
\ee
We will show \eqref{uni3} to hold for every even-odd combination of $(m,m')$.

For $(m,m') \in (\mbox{odd},\mbox{odd})$, from Eqs.~\eqref{alpha_beta} and \eqref{rho_sigma}, we obtain 
\be
	I^{(\gamma)}_{ mm^\prime }
	=
	\frac{ m+m' }{ \sqrt{ mm' } \pi^2 }
	K_{mm'},
\;\;\;
	J^{(\gamma)}_{ mm^\prime }
	=
	- \frac{ m-m' }{ \sqrt{ mm' } \pi^2 }
	K_{mm'},
\label{IJ2}
\ee
where
\begin{align}
	K_{mm'}
	&:=
	\sum_{n=1}^\infty
	\frac{ n^2 }{ [ n^2-(m/2)^2 ][ n^2-( m'/2 )^2 ] }
\nn
\\
	&=
	\sum_{n=1}^\infty \frac{1}{n^2-(m'/2)^2}
	+
	(\frac{m}{2})^2 \sum_{n=1}^\infty \frac{1}{ [ n^2-(m/2)^2 ][ n^2-( m'/2 )^2 ] }.
\label{K}
\end{align}
The summations over $n$ in Eq.~\eqref{K} are calculated to give
\be
	K_{ mm^\prime } = \frac{\pi^2}{4} \delta_{mm^\prime},
\label{K2}
\ee
using the following formulas~\cite[pp.~68--69]{iwa}
\begin{align}
	\sum_{k=1}^\infty \frac{1}{y^2-k^2}
	&=
	\frac{\pi}{2y} \cot (\pi y) -\frac{1}{2y^2},
\\
	\sum_{k=1}^\infty \frac{1}{ [ (ky)^2-1]^2}
	&=
	\frac{\pi^2}{4y^2 } {\rm cosec}^2 ( \frac{\pi}{y} ) 
	+
	\frac{\pi}{4y} \cot ( \frac{\pi}{y} ) - \frac12.
\end{align}
Substituting Eq.~\eqref{K2} into Eq.~\eqref{IJ2}, we see Eq.~\eqref{uni3} to hold in this case.

For $(m,m') \in (\mbox{odd},\mbox{even})$, using Eqs.~\eqref{alpha_beta} and \eqref{rho_sigma} again, we obtain
\be
	I^{(\gamma)}_{mm'}
	=
	- \frac{ (-1)^{\gamma-1} }{ (m-m')\pi } \sqrt{ \frac{m'}{m} },
\;\;\;
	J^{(\gamma)}_{mm'}
	=
	\frac{ (-1)^{\gamma-1} }{ (m+m')\pi } \sqrt{ \frac{m'}{m} }.
\label{IJ3}
\ee
From this, we see Eq.~\eqref{uni3} to hold in this case.

Finally, for $(m,m') \in (\mbox{even},\mbox{even})$, using Eqs.~\eqref{alpha_beta} and \eqref{rho_sigma} again, we obtain
\be
	I^{(\gamma)}_{mm'}
	=
	\frac12 \delta_{mm^\prime},
\;\;\;
	J^{(\gamma)}_{mm'}
	=
	0.
\ee
From this, we see Eq.~\eqref{uni3} to hold in this case.

\section{Green-function method}
\label{sec:green}

We show another derivation of the vacuum expectation values of energy-momentum tensor, Eqs.~\eqref{Tab_02}, \eqref{Tab_03}, \eqref{Tab_12}, and \eqref{Tab_13} by the Green-function method (see, e.g., \cite{Birrell:1982ix}).
 
  \subsection{Green functions}

The mode function $g_m$, Eq.~\eqref{g}, is rewritten as 
\begin{align}
   g_m (z_-,z_+) = 
    \left\{
     \begin{aligned}
      \frac{1}{2\sqrt{m\pi}}& 
      \left(
      e^{-i\frac{m\pi}{L}z_-} + e^{-i\frac{m\pi}{L}z_+}
      \right) &\quad (m &: \text{odd}) \\
      \frac{1}{2i\sqrt{m\pi}}& 
      \left(
      e^{-i\frac{m\pi}{L}z_-} - e^{-i\frac{m\pi}{L}z_+}
      \right) &\quad (m &: \text{even})
     \end{aligned}
    \right. .
  \end{align}
Then, Hadamard's elementary function is computed as
\begin{align}
    & \bar{G} (z_-,z_+;z'_-,z'_+)  :=
    \langle 0_g | \{\phi(z_-,z_+), \phi(z'_-,z'_+)\} |0_g\rangle   
 \label{eq:Hadamard_g} \\
    &= \sum_{m=1}^\infty [g_m(z_-,z_+) g_m^*(z'_-,z'_+)
    + g_m^*(z_-,z_+) g_m(z'_-,z'_+)]
    \\
    &= \frac{1}{4\pi} \sum_{m=1}^\infty \frac{1}{m} 
    \left[
    e^{-i \frac{m\pi}{L} \Delta z_-} + e^{-i \frac{m\pi}{L} \Delta z_+}
    - e^{-i \frac{m\pi}{L} (z_- - z'_+ +L)}
    - e^{-i \frac{m\pi}{L} (z_+ - z'_- +L)}
    \right] + \text{c.c.} \\
    &= 
    - \frac{1}{4\pi} \ln 
    \left[
    \frac{\sin^2(\pi \Delta z_- /2L) \sin^2(\pi \Delta z_+ /2L)}
    {\cos^2(\pi (z_- - z'_+) /2L) \cos^2(\pi (z_+ - z'_-)/2L)}
    \right] ,
\end{align} 
  where $\Delta z_\pm := z_\pm - z'_\pm$ and ${\rm c.c.}$ denotes the complex conjugate.
  In the last line, we have performed the summation over $m$ after replacement
  $\Delta z_\pm \to \Delta z_\pm - i\epsilon$, where $\epsilon$ is a real small parameter, in order to make it converge.
  On the other hand, the Pauli--Jordan function is 
  \begin{align}
    & i G (z_-,z_+;z'_-,z'_+) :=
    \langle 0_g | [\phi(z_-,z_+), \phi(z'_-,z'_+)] |0_g\rangle \\
    &= \sum_{m=1}^\infty [g_m(z_-,z_+) g_m^*(z'_-,z'_+)
    - g_m^*(z_-,z_+) g_m(z'_-,z'_+)]
    \\
    &= \frac{1}{4\pi} \sum_{m=1}^\infty \frac{1}{m} 
    \left[
    e^{-i \frac{m\pi}{L} \Delta z_-}
    + e^{-i \frac{m\pi}{L} \Delta z_+}
    - e^{-i \frac{m\pi}{L} (z_- - z'_+ +L)}
    - e^{-i \frac{m\pi}{L} (z_+ - z'_- +L)}
    \right] - \text{c.c.} \\
    &= 
    - \frac{i}{2}
    \sum_{m=-\infty}^{\infty}
    [\theta (\Delta z_- - 2mL) + \theta (\Delta z_+ - 2mL)
    - \theta (z_- - z'_+ -(2m-1)L) - \theta (z_+ - z'_- -(2m-1)L)] ,
   \label{eq:PJ_g}
  \end{align}  
where $\theta$ denotes the step function. Here, we have used the following formulas,
\begin{align}
   \sum_{n=1}^\infty \frac{e^{-in(x-i\epsilon)} }{n} 
    = - \ln [1 - e^{-i(x-i\epsilon)}] , \;
   \ln (-x+i\epsilon) - \ln (-x-i\epsilon) = 2\pi i \theta(x) , \;
   \frac{ \sin (\pi x) }{\pi x}  =  \prod_{n=1}^\infty \left(1-\frac{x^2}{n^2}\right) .
\end{align}
  
The mode functions $f_n^{(\gamma)}$, Eq.~\eqref{f}, on each support is rewritten as 
  \begin{equation}
   \left\{
    \begin{aligned}
     f^{(1)}(z_-, z_+) =& \frac{1}{2i\sqrt{n\pi}}
     \left(
     e^{-i\frac{2n\pi}{L}z_-} - e^{-i\frac{2n\pi}{L}z_+}
     \right) &\quad &(0 \le x \le L/2) \\
     f^{(2)}(z_-, z_+) =& - \frac{1}{2i\sqrt{n\pi}} 
     \left(
     e^{-i\frac{2n\pi}{L}z_-} - e^{-i\frac{2n\pi}{L}z_+}
     \right) &\quad &(-L/2 \le x \le 0)
    \end{aligned}
   \right. .
  \end{equation}
Hadamard's elementary function and Pauli-Jordan functions are given by
   \begin{align}
    & \bar{F}(z_-,z_+;z'_-,z'_+) :=
   \langle 0_f | \{\phi(z_-,z_+), \phi(z'_-,z'_+)\} |0_f\rangle 
   =
   \sum_{\gamma=1}^2 \bar{F}^{(\gamma)}(z_-,z_+;z'_-,z'_+),
\\
	& i F (z_-,z_+;z'_-,z'_+) :=
    \langle 0_f | [\phi(z_-,z_+), \phi(z'_-,z'_+)] |0_f\rangle 
    =
	\sum_{\gamma=1}^2 i F^{(\gamma)} (z_-,z_+;z'_-,z'_+),
\end{align}
where
\begin{align}
	&\bar{F}^{(\gamma)}(z_-,z_+;z'_-,z'_+)
    := \sum_{n=1}^\infty [f^{(\gamma)}_n(z_-,z_+) f_n^{(\gamma) \ast}(z'_-,z'_+)
    + f_n^{(\gamma ) \ast}(z_-,z_+) f_n^{( \gamma )}(z'_-,z'_+)]
    \\
    &= \frac{1}{4\pi} \sum_{n=1}^\infty \frac{1}{n} 
    \left[
    e^{-i \frac{2n\pi}{L} \Delta z_-} + e^{-i \frac{2n\pi}{L} \Delta z_+}
    - e^{-i \frac{2n\pi}{L} (z_- - z'_+)}
    - e^{-i \frac{2n\pi}{L} (z_+ - z'_-)}
    \right] + \text{c.c.}
\label{Fbar1} \\
    &= 
    - \frac{1}{4\pi} \ln 
    \left[
    \frac{\sin^2(\pi \Delta z_-/L) \sin^2(\pi \Delta z_+/L)}
    {\sin^2(\pi (z_- - z'_+) /L) \sin^2(\pi (z_+ - z'_-)/L)}
    \right],
\label{Fbar2}
    \\
	&i F^{(\gamma)} (z_-,z_+;z'_-,z'_+)
    := \sum_{n=1}^\infty [f_n^{(\gamma)} (z_-,z_+) f_n^{(\gamma) \ast}(z'_-,z'_+)
    - f_n^{(\gamma) \ast}(z_-,z_+) f_n^{(\gamma)}(z'_-,z'_+)]
    \\
    &= \frac{1}{4\pi} \sum_{n=1}^\infty \frac{1}{n} 
    \left[
    e^{-i \frac{2n\pi}{L} \Delta z_-} + e^{-i \frac{2n\pi}{L} \Delta z_+}
    - e^{-i \frac{2n\pi}{L} (z_- - z'_+)}
    - e^{-i \frac{2n\pi}{L} (z_+ - z'_-)}
    \right] - \text{c.c.}
\label{F1}
\\
    &=
    - \frac{i}{2}
    \sum_{n=-\infty}^{\infty}
    [\theta (\Delta z_- - nL) + \theta (\Delta z_+ - nL)
    - \theta (z_- - z'_+ -nL) - \theta (z_+ - z'_- -nL)] .
\label{F2}
\end{align}
Hereafter, we should keep in mind that $\bar{ F }^{(\gamma)}$ and $F^{(\gamma)}$ are non-zero and given by the above expressions, i.e., Eqs.~\eqref{Fbar1}, \eqref{Fbar2}, \eqref{F1}, and \eqref{F2}, only on each support of $f^{(\gamma)}$ ($\gamma=1,2$). For example, $\bar{ F }^{(1)}=0$ in $-L/2 \leq x \leq 0$ and $\bar{F}^{(2)}=0$ in $0 \leq x \leq L/2$.

  \subsection{Appearance of the Dirichlet wall}

  For $t<0$, the energy-momentum tensor is given by    
   \begin{align}
    &\langle 0_g | {\bm T}_{\pm\pm} | 0_g \rangle_{t<0}
	=
    \frac{1}{2}\lim_{z'_\pm\to z_\pm}
    \partial_\pm \partial'_\pm \bar{G} (z_-,z_+;z'_-,z'_+) \\
    &= \frac{\pi}{8L^2} \lim_{\Delta z_\pm\to 0} \sum_{m=1}^\infty m 
    \left(
    e^{-i\frac{m\pi}{L}\Delta z_\pm}
    + e^{i\frac{m\pi}{L}\Delta z_\pm}
    \right) \\
    &= - \frac{\pi}{16L^2} \lim_{\Delta z_\pm\to 0}
    \frac{1}{\sin^2(\pi\Delta z_\pm /2L)} 
    = - \frac{\pi}{48L^2}
    - \lim_{\Delta z_\pm\to 0} \frac{1}{4\pi \Delta z_\pm^2} ,
   \end{align}
  where the last divergent term, which is
  independent of the cavity size $L$, can be subtracted as the
  zero-point energy with the cavity size $L = \infty$. Thus, we have reproduced Eq.~\eqref{Tab_02} by the Green-function method.

  For $t>0$, the energy-momentum tensor is given by    
  \begin{equation}
    \langle 0_g | {\bm T}_{\pm\pm} | 0_g \rangle_{t>0} =
    \frac{1}{2}\lim_{A\to B}
    \partial_\pm \partial'_\pm [(i F_{AC}) (i F_{BD}) \bar{G}_{CD}] ,
  \end{equation}
  where 
  $A:=(z_-,z_+)$ and $B:=(z'_-,z'_+)$.
  In abbreviated notation, a capital Latin index denotes one point on the
  spacetime and the same indices denote the Klein-Gordon product, 
  such as $F_{AB} = F(z_-,z_+;z'_-,z'_+)$ and 
  $\phi_A \psi_A = \langle \phi, \psi \rangle$.

  For $t>0$, we can also obtain the energy-momentum tensor from a Green
  function.
  However, 
  since the boundary condition has changed for $t>0$, the Green function
  for $t>0$ will differ from one defined by (\ref{eq:Hadamard_g}).
  In order to obtain the Green function for $t>0$, we should
  propagate $\bar{G} $ at $t=0$, by using $F$.
  For example, if a scalar field is given by $\phi_0(z_-,z_+)$ and 
  $\partial_t\phi_0(z_-,z_+)$ at $t=0$ as initial data, then for $t>0$
  we have  
  \begin{equation}
    \phi_{t>0}(z_-,z_+) 
    = \langle i F(z_-,z_+;z'_-,z'_+) , 
    \phi_0(z'_-,z'_+) \rangle |_{t'=0}\\
    = \left.\int dx' 
    (F \partial_{t'} \phi_{0} - \phi_0 \partial_{t'} F) 
    \right|_{t'=0} .
  \end{equation}

Using 
   \begin{align}
     \partial_\pm F(z_-,z_+;z'_-,z'_+) &=
    - \frac{1}{2}\sum_{n=-\infty}^{\infty}
   [\delta(\Delta z_\pm - nL) - \delta(z_\pm - z'_\mp - nL)] , \\
     \partial_\pm \partial_{t'} F(z_-,z_+;z'_-,z'_+) 
	&=
    \mp \frac{1}{2}\sum_{n=-\infty}^{\infty}
   \partial_{x'} [\delta(\Delta z_\pm - nL) + \delta(z_\pm - z'_\mp -nL)] ,
   \end{align}
we have for $0\le x \le L/2$,
   \begin{align}
    & \partial_\pm \langle i F^{(1)} (z_-,z_+;z'_-,z'_+), 
    g_m(z'_-,z'_+) \rangle |_{t'=0} 
   = \int_0^{L/2}\!\!\! dx'
    [\partial_\pm F^{(1)} 
    \partial_{t'} g_m - g_m\partial_\pm \partial_{t'} F^{(1)} ] \\
    &= - \sum_{n=-\infty}^\infty \int^{L/2}_0 \!\!\! dx'
    [\delta (z_\pm \mp x' - nL)\partial'_\pm g_m
    - \delta (z_\pm \pm x' - nL)\partial'_\mp g_m ]
\nonumber
\\
  &
	\pm \left.\frac{1}{2}\sum_{n=-\infty}^\infty
    [\delta (z_\pm \mp x' - nL) + \delta (z_\pm \pm x' - nL)]g_m
    \right|^{x'=L/2}_{x'=0}  .
   \end{align}
We evaluate them respectively as
   \begin{align}
&    \partial_+ \langle i F^{(1)}, 
    g_m \rangle |_{t'=0}
   = - \sum_{n=-\infty}^\infty \int^{L/2}_0 \!\!\! dx'
    [\delta (z_+ - x' - nL)\partial'_+ g_m
    - \delta (z_+ + x' - nL)\partial'_- g_m ] 
\nonumber
\\
    &
    + \left.\frac{1}{2}\sum_{n=-\infty}^\infty
    [\delta (z_+ - x' - nL) + \delta (z_+ + x' - nL)]g_m
    \right|^{x'=L/2}_{x'=0}  \\
    &= - \sum_{n=-\infty}^\infty 
    [
    \Pi_0^{L/2}(z_+ - nL)\partial'_+ g_m|_{x' = z_+ - nL}
    - 
    \Pi_{-L/2}^{0}(z_+ - nL)\partial'_- g_m|_{x' = - z_+ + nL} ]
    - \sum_{n=-\infty}^\infty
    \delta (z_+ - nL) g_m|_{x'=0}\\
    &=
    \begin{cases}
    \displaystyle
     -  \sum_{n=-\infty}^\infty 
     \left\{
     \frac{\delta (z_+ - nL)}{\sqrt{m\pi}}
     + i\frac{\sqrt{m\pi}}{2L} e^{-i\frac{m\pi}{L}z_+}
     (-1)^n[\Pi_{-L/2}^0(z_+ - nL) - \Pi_{0}^{L/2}(z_+ - nL)]
     \right\} 
     &
     (m:{\rm odd})
	\\
	\displaystyle
     - \frac{\sqrt{m\pi}}{2L} e^{-i\frac{m\pi}{L}z_+}
     \sum_{n=-\infty}^\infty
     \Pi_{-L/2}^{L/2}(z_+ - nL)
     &
     (m:{\rm even})
    \end{cases},
 \end{align}
and
   \begin{align}
&    \partial_- \langle i F^{(1)} , 
    g_m \rangle |_{t'=0}
    = - \sum_{n=-\infty}^\infty \int^{L/2}_0 \!\!\! dx'
    [\delta (z_- + x' - nL)\partial'_- g_m
    - \delta (z_- - x' - nL)\partial'_+ g_m ] \nonumber \\
    &- \left.\frac{1}{2}\sum_{n=-\infty}^\infty
    [\delta (z_- + x' - nL) + \delta (z_- - x' - nL)]g_m
    \right|^{x'=L/2}_{x'=0} \\
    & = \sum_{n=-\infty}^\infty 
    [
    \Pi_0^{L/2}(z_- - nL)\partial'_+ g_m|_{x' = z_- - nL}
    - 
    \Pi_{-L/2}^{0}(z_- - nL)\partial'_- g_m|_{x' = - z_- + nL} ]
    + \sum_{n=-\infty}^\infty
    \delta (z_- - nL) g_m|_{x'=0}\\
    &=
    \begin{cases}
    \displaystyle
     \sum_{n=-\infty}^\infty 
     \left\{
     \frac{\delta (z_- - nL)}{\sqrt{m\pi}}
     + i\frac{\sqrt{m\pi}}{2L} e^{-i\frac{m\pi}{L}z_-}
     (-1)^n[\Pi_{-L/2}^0(z_- - nL) - \Pi_{0}^{L/2}(z_- - nL)]
     \right\}
	& (m:{\rm odd})	\\
      \displaystyle
     \frac{\sqrt{m\pi}}{2L} e^{-i\frac{m\pi}{L}z_-}
     \sum_{n=-\infty}^\infty
     \Pi_{-L/2}^{L/2}(z_- - nL)
     & (m:{\rm even})
    \end{cases}.
   \end{align}
The last two results of calculation are written in the following short form
  \begin{equation}
   \begin{aligned}
    &\partial_\pm \langle i F^{(1)} (z_-,z_+;z'_-,z'_+), 
    g_m(z'_-,z'_+) \rangle |_{t'=0}\\
    &=
    \left\{
    \begin{aligned}
     &\mp  \sum_{n=-\infty}^\infty 
     \left[
     \frac{1}{\sqrt{m\pi}}\delta (z_\pm - nL)
     + i\frac{\sqrt{m\pi}}{2L} e^{-i\frac{m\pi}{L}z_\pm}
     (-1)^{n+1} \Pi_0^L (z_\pm - nL)
     \right] & &(m:\text{odd})\\
     &\mp \frac{\sqrt{m\pi}}{2L} e^{-i\frac{m\pi}{L}z_\pm}
     & &(m:\text{even})
    \end{aligned}
    \right. .
   \end{aligned}
  \end{equation}
  Here, $\Pi_a^b(x)$ ($a<b$) denotes a rectangle function defined by 
  \begin{equation}
   \Pi_a^b(x) :=
    \int^b_a \delta(x-y) dy= 
    \left\{
     \begin{aligned}
      &0 & \quad &(x>b, x<a)\\
      &1 & \quad &(a<x<b) \\
      &\frac{1}{2} & \quad &(x=a,b)
     \end{aligned}
    \right. .
  \end{equation}
  
  As a result, for $t>0$ the energy-momentum tensor is given by 
  \begin{equation}
   \begin{aligned}
    \langle 0_g | {\bm T}_{\pm\pm} | 0_g \rangle_{t>0}
    =& \frac{1}{2}\lim_{\Delta z_\pm\to 0}
    \sum_{m=1}^\infty
    \partial_\pm \langle i F^{(1)}(z_-,z_+;U,V), g_m(U,V) \rangle 
    \partial'_\pm \langle i F^{(1)}(z'_-,z'_+;U',V'), g_m(U',V') \rangle^* + \text{c.c.}\\
    =& 
    \left\{
    \begin{aligned}
     &
     \sum_{\substack{m=1\\m:\text{odd}}}^\infty
     \frac{1}{m\pi} \delta(z_\pm - nL)^2
     + \frac{1}{2} \lim_{\Delta z_\pm \to 0}
     \sum_{\substack{m=2\\m:\text{even}}}^\infty 
     \frac{m\pi}{4L^2}
     \left(e^{-i\frac{m\pi}{L}\Delta z_\pm}
     + e^{i\frac{m\pi}{L}\Delta z_\pm}\right)
     & \quad (z_\pm &= n L) \\
     &\frac{1}{2} \lim_{\Delta z_\pm \to 0}
     \sum_{m=1}^\infty \frac{m\pi}{4L^2}
     \left(e^{-i\frac{m\pi}{L}\Delta z_\pm}
     + e^{i\frac{m\pi}{L}\Delta z_\pm}\right)
     & \quad (z_\pm &\neq n L) \\
    \end{aligned}
    \right. 
    \\
    =& 
    \left\{
    \begin{aligned}
     &
     \sum_{\substack{m=1\\m:\text{odd}}}^\infty
     \frac{1}{m\pi} \delta(z_\pm - nL)^2
     - \frac{\pi}{24L^2}
     - \lim_{\Delta z_\pm \to 0} \frac{1}{8\pi\Delta z_\pm^2}
     & \quad (z_\pm &= n L) \\    
     & - \frac{\pi}{48L^2} 
     - \lim_{\Delta z_\pm \to 0} \frac{1}{4\pi\Delta z_\pm^2}
     & \quad (z_\pm &\neq n L) \\
    \end{aligned}
    \right. 
    ,
   \end{aligned}
   \label{eq:energy_momentum_g}
  \end{equation}
  where $n$ is a non-negative integer. 
  Note that, while we have focused on the domain $0\le x \le L/2$ using
  $F^{(1)}$, the energy-momentum tensor as well as
  (\ref{eq:energy_momentum_g}) can be obtained by using $F^{(2)}$
  also for $-L/2\le x \le 0$. Thus, we reproduce Eq.~\eqref{Tab_03} by the Green-function method.

We can implement the regularization of the energy-momentum tensor following the standard procedure. We subtract the divergent term which exists even in the flat spacetime with Dirichlet walls at $x=\pm L/2$ in the limit $L\to \infty$. Then, the renormalized energy-momentum tensor is then given by
\begin{eqnarray}
    \langle 0_g | {\bm T}_{\pm\pm} | 0_g \rangle_{{\rm ren}, t<0}&=& 
-\frac{\pi}{48L^{2}}, \\
    \langle 0_g | {\bm T}_{\pm\pm} | 0_g \rangle_{{\rm ren}, t>0}&=& 
    \left\{
    \begin{aligned}
     &
     \sum_{\substack{m=1\\m:\text{odd}}}^\infty
     \frac{1}{m\pi} \delta(z_\pm - nL)^2
     - \frac{\pi}{24L^2}
     + \lim_{\Delta z_\pm \to 0} \frac{1}{8\pi\Delta z_\pm^2}
     &  (z_\pm &= n L) \\
     & - \frac{\pi}{48L^2} 
     &  (z_\pm &\neq n L) \\
    \end{aligned}
    \right. .
\label{Tab_13_ren}
\end{eqnarray}
We can see that the energy-momentum tensor is still divergent on the null lines for $t>0$. 

  \subsection{Disappearance of the Dirichlet wall}

  For $t<0$, the energy-momentum tensor is given by 
   \begin{align}
&    \langle 0_f | {\bm T}_{\pm\pm} | 0_f \rangle_{t<0} =
    \frac{1}{2}\lim_{z'_\pm \to z_\pm}
    \partial_\pm \partial'_{\pm} \bar{F} (z_-,z_+;z'_-,z'_+)\\
    &= \frac{\pi}{2L^2} \lim_{\Delta z_\pm \to 0}
    \sum_{n=1}^\infty n 
    \left(e^{-i\frac{2n\pi}{L}\Delta z_\pm}
    + e^{i\frac{2n\pi}{L}\Delta z_\pm}
    \right) 
    = - \frac{\pi}{12L^2} - \lim_{\Delta z_\pm \to 0} 
    \frac{1}{4\pi \Delta z_\pm^2}.
   \end{align}
Thus, we reproduce Eq.~\eqref{Tab_12} after subtracting the last diverging term.

For $t>0$, the energy-momentum tensor is given by    
  \begin{equation}
   \begin{aligned}
    \langle 0_f | {\bm T}_{\pm\pm} | 0_f \rangle_{t>0} =&
    \frac{1}{2}\lim_{A\to B}
    \partial_\pm \partial'_{\pm} [(i G_{AC}) (iG_{BD}) \bar{F}_{CD}],
   \end{aligned}
  \end{equation}
where $A=(z_-,z_+)$ and $B=(z'_-,z'_+)$ again.

  Using 
   \begin{align}
    \partial_\pm G(z_-,z_+;z'_-,z'_+) =&
    - \frac{1}{2}\sum_{m=-\infty}^{\infty}
   [\delta(\Delta z_\pm - 2mL) - \delta(z_\pm - z'_\mp - (2m-1)L)] ,\\
    \partial_\pm \partial_{t'} G(z_-,z_+;z'_-,z'_+) =&
    \mp \frac{1}{2}\sum_{m=-\infty}^{\infty}
   \partial_{x'} [\delta(\Delta z_\pm - 2mL)
    + \delta(z_\pm - z'_\mp - (2m-1)L)] ,
   \end{align}
  we have 
   \begin{align}
&    \partial_\pm \langle iG(z_-,z_+;z'_-,z'_+), 
    f^{(1)}_n(z'_-,z'_+) \rangle |_{t'=0}
    = \int_0^{L/2}\!\!\! dx'
    [\partial_\pm G \partial_{t'} f^{(1)}_n
    - f^{(1)}_n\partial_\pm \partial_{t'} G] \\
    & = - \sum_{m=-\infty}^\infty
    \int^{L/2}_0 \!\!\! dx'
    \left[
    \delta (z_\pm \mp x' - 2mL) \partial'_\pm f^{(1)}_n
    - \delta (z_\pm \pm x' - (2m-1)L) \partial'_\mp f^{(1)}_n 
    \right]
\nonumber    \\
    & \pm \left.\frac{1}{2}\sum_{m=-\infty}^\infty
    [\delta (z_\pm \mp x' -2mL)
    + \delta (z_\pm \pm x' - (2m-1)L)]f^{(1)}_n
    \right|^{x'=L/2}_{x'=0} \\
    &= \mp \frac{\sqrt{n\pi}}{L} e^{-i\frac{2n\pi}{L}z_\pm}
    \sum_{m=-\infty}^\infty \Pi_0^L (\pm z_\pm - 2mL)
   \end{align}
  and 
   \begin{align}
    \partial_\pm \langle iG(z_-,z_+;z'_-,z'_+), 
    f^{(2)}_n(z'_-,z'_+) \rangle |_{t'=0}
    =& \int_{-L/2}^{0}\!\!\! dx'
    [\partial_\pm G \partial_{t'} f^{(2)}_n
    - f^{(2)}_n\partial_\pm \partial_{t'} G] \\
    =& \pm \frac{\sqrt{n\pi}}{L} e^{-i\frac{2n\pi}{L}z_\pm}
    \sum_{m=-\infty}^\infty \Pi_{-L}^0 (\pm z_\pm - 2mL) .
   \end{align}
  
  As a result, for $t>0$ the energy-momentum tensor is given by 
   \begin{align}
   &
    \langle 0_f | {\bm T}_{\pm\pm} | 0_f \rangle_{t>0} \nonumber \\
&=
    \frac{1}{2} \lim_{\Delta z_\pm \to 0}
    \sum_{\gamma = 1}^2 \sum_{n=1}^\infty 
    \partial_\pm \langle iG(z_-,z_+;U,V), f^{(\gamma)}_n(U,V) \rangle
    \partial'_\pm \langle iG(z'_-,z'_+;U',V'), f^{(\gamma)}_n(U',V') \rangle^*
    + \text{c.c.}  \\
    &=
    \left\{
    \begin{aligned}
     &\frac{1}{4} \lim_{\Delta z_\pm \to 0}\sum_{n=1}^\infty
     \frac{n\pi}{L^2} 
     (e^{-i\frac{2n\pi}{L}\Delta z_\pm}
     + e^{i\frac{2n\pi}{L}\Delta z_\pm})
     & \quad &(z_\pm = m L) \\
     &\frac{1}{2} \lim_{\Delta z_\pm \to 0}\sum_{n=1}^\infty
     \frac{n\pi}{L^2} 
     (e^{-i\frac{2n\pi}{L}\Delta z_\pm}
     + e^{i\frac{2n\pi}{L}\Delta z_\pm})
     & \quad &(z_\pm \neq m L)
     \\
    \end{aligned}
    \right. 
    \\
    &=
    \left\{
    \begin{aligned}
     & - \frac{\pi}{24L^2}
     - \lim_{\Delta z_\pm \to 0} \frac{1}{8\pi \Delta z_\pm^2}
     & \quad &(z_\pm = m L) \\
     & - \frac{\pi}{12L^2}
     - \lim_{\Delta z_\pm \to 0} \frac{1}{4\pi \Delta z_\pm^2}
     & \quad &(z_\pm \neq m L)
     \\
    \end{aligned}
    \right.  ,
   \end{align}
where $m$ is a non-negative integer. Thus, we have reproduced Eq.~\eqref{Tab_13} with the Green-function method. The renormalized energy-momentum tensor is given by 
\begin{eqnarray}
    \langle 0_f | {\bm T}_{\pm\pm} | 0_f \rangle_{{\rm ren}, t<0}&=& 
-\frac{\pi}{12L^{2}}, \\
    \langle 0_f | {\bm T}_{\pm\pm} | 0_f \rangle_{{\rm ren}, t>0}&=& 
    \left\{
    \begin{aligned}
     &     - \frac{\pi}{24L^2}
     + \lim_{\Delta z_\pm \to 0} \frac{1}{8\pi\Delta z_\pm^2}
     & \quad (z_\pm &= n L) \\
     & - \frac{\pi}{12L^2} 
     & \quad (z_\pm &\neq n L) \\
    \end{aligned}
    \right. .
\end{eqnarray}
We can see that the energy-momentum tensor is still divergent on the null lines for $t>0$.


\end{document}